\documentclass[twocolumn,showpacs,preprintnumbers,amsmath,amssym]{revtex4}

\usepackage{graphicx}
\usepackage{dcolumn}
\usepackage{bm}

\newcommand{\eps}[1] {\varepsilon_{c}^{({#1})}}

\begin{document}

\title{Harmonic {\it vs.} subharmonic patterns in a spatially
 forced oscillating chemical reaction}

\author{Martin  Hammele$^{1,2}$ and Walter Zimmermann$^2$}
\affiliation{%
$^1$Theoretische Physik, Universit\"at Bayreuth, 95440 Bayreuth, Germany\\
$^2$Theoretische Physik, Universit\"at des Saarlandes, 66041 Saarbr\"ucken, Germany
}

\date{\today}

\begin{abstract}
The effects of a spatially periodic forcing on an oscillating 
chemical reaction as described by the Lengyel-Epstein model 
are investigated. We find a surprising competition between 
two oscillating patterns,  where one  is harmonic and the other  subharmonic 
with respect to the spatially periodic forcing.  The occurrence 
of a subharmonic pattern is remarkable as well  as its 
preference up to rather large values of the modulation amplitude. For 
small modulation amplitudes we derive from the model system 
a generic equation for the envelope of the oscillating reaction 
that includes an additional forcing contribution, 
compared to the amplitude equations known from 
previous studies in other systems. 
The analysis of this amplitude equation allows 
the derivation of analytical expressions even for the forcing 
corrections to the threshold and to the oscillation frequency,
which are in a wide range of parameters in good  agreement 
with the numerical analysis of the complete reaction 
equations. In the nonlinear regime beyond threshold,  the 
subharmonic solutions exist in a finite range of the control 
parameter that has been determined by solving the
reaction equations numerically for various sets of
parameters.
\end{abstract}


\maketitle
\section{\label{intro}Introduction}

Studying the response of pattern-forming systems with respect 
to an external stimulus  provides a powerful method
to investigate  the inherently nonlinear mechanism of
self-organization in various systems under nonequilibrium conditions.
Thermal convection  \cite{Kelly:78.1} and 
electroconvection in nematic liquid crystals \cite{Lowe:83.1,Lowe:85.1,Lowe:86.1} 
are two  systems, 
where the effects of spatially periodic forcing
have been investigated rather early. 
Further on, the effects of forcing on stationary
bifurcations 
have been studied extensively in many different systems 
\cite{Coullet:86.2,Coullet:87.1,Zimmermann:93.3,Zimmermann:96.1,Zimmermann:96.2,Zimmermann:96.3,Zimmermann:1998.1,Epstein:2001.2,Zimmermann:2004.1}, 
and this branch of nonlinear science has also evolved to  
forcing studies on oscillatory media and traveling waves 
\cite{Riecke:88.1,Walgraef:88.1,Rehberg:88.1,Coullet:89.1,Coullet:92.3,Coullet:92.4,Coullet:92.5,Walgraef:96.1,Meron:1998.1,Swinney:97.1,Swinney:2000.1}.

Recently, the response behavior of  patterns with respect to a 
combination of spatial and temporal forcing 
has attracted a great deal of attention because of 
the development of  flexible forcing techniques 
using illumination,  as for instance in 
photosensitive chemical reactions 
\cite{Swinney:97.1,Swinney:2000.1,Zimmermann:2002.2,Epstein:2003.1,Sagues:2003.1,Sagues:2004.1} 
or in electroconvection
in nematic liquid crystals \cite{Henriot:2003.1,Schuler:2004.1}. 
For the  photosensitive
chlorine dioxide-iodine-malonic acid (CDIMA) reaction, 
as described by the so-called Lengyel-Epstein 
model \cite{Lengyel:1991.1,Lengyel:1993.1}, one finds
in a large parameter range Turing patterns. In particular,
their response   with respect to a forcing of a traveling wave type, 
which is spatially resonant or near-resonant
with respect to the characteristic  wavelength of the
Turing pattern, exhibits a number of new phenomena 
and has therefore attracted considerable attention
recently \cite{Sagues:2004.1,Sagues:2003.1}.

The Lengyel-Epstein model also exhibits  a spatially 
homogeneous and supercritical Hopf bifurcation 
\cite{Lengyel:1991.1,Epstein:1999.1,Epstein:1999.3}
similar to the one  found in other chemical reactions. 
In the present work,
we investigate the response of the
Hopf bifurcation of this model with respect to
a spatially periodic but time-independent illuminating 
forcing,  which enters the   Lengyel-Epstein model additively. 
Beyond the instability of the homogeneous
chemical reaction, we find a surprising competition between a
temporally oscillating and spatially  modulated 
reaction that is harmonic with
respect to the spatially periodic forcing 
and another one which is subharmonic. Besides its occurrence close to
the threshold of the Hopf bifurcation,  also
the preference of the subharmonic pattern  up to large 
forcing amplitudes is remarkable as well.

From a technical point of view, our
 analysis  is related to  a previous study of a complex
Ginzburg-Landau equation corresponding to a
spatially periodic  modulation of a temporally resonant
forcing of a chemical reaction \cite{Zimmermann:2002.2}. 
While the forcing
in the previous studies entered the model equation
multiplicatively, the forcing enters the Lengyel-Epstein model
additively. Close to the threshold of the Hopf bifurcation,
we also reduced  the Lengyel-Epstein model to
a universal equation for the amplitude of the oscillations
by using a multiple scale perturbation
technique \cite{CrossHo}.
We find that the forcing contribution occurs also   
multiplicatively in the resulting amplitude equation, 
but the forcing contribution to
the complex amplitude equation has a different form,
compared to previous studies. Nevertheless,  as a result
of the analysis of this  amplitude 
equation,  we also obtain a competition between
harmonic and subharmonic structures that agree  in
the limit of small modulation amplitudes very well with the 
results of
the full numerical analysis of the Lengyel-Epstein model.

This work is organized as follows.
In Sec.~\ref{sec:2} we present the Lengyel-Epstein model
for a spatially modulated illumination and
in  Sec.~\ref{sec:3} 
we determine its stationary basic states and study
their stability against small 
perturbations for a uniform
and spatially modulated illumination. 
Some   results of numerical simulations of the full
nonlinear model equations
are presented in Sec.~\ref{sec:4}.
Close to threshold, in the so-called weakly nonlinear regime,
the dynamical behavior of 
the Lengyel-Epstein model can be described
in terms of an  amplitude equation as discussed  in Sec.~\ref{sec:5}.  
The effect  of the modulation on the threshold
of this amplitude equation 
is investigated in Sec.~\ref{sec:5.1}
by using two different approaches   given by
a perturbation calculation and 
by a fully numerical solution of the general linear problem. 
The results are discussed  in detail in Sec.~\ref{sec:5.2},
where we also make a comparison with the thresholds obtained from 
a direct solution of the Lengyel-Epstein model.
The work  is finished 
with a summary and some concluding remarks in Sec.~\ref{sec:6}.
A detailed derivation of the amplitude equation from the 
Lengyel-Epstein model is given in the Appendix.

\section{The Lengyel-Epstein model}
\label{sec:2}

The starting point of  our investigations 
on the effects of a spatially periodic modulated control
parameter on a chemical reaction is 
the Lengyel-Epstein model \cite{Lengyel:1991.1,Lengyel:1993.1}.
This model describes two different  instabilities
of a spatially homogeneous   chemical reaction, being  either 
a Turing instability to a 
stationary and  spatially periodic pattern
or a 
Hopf bifurcation to a spatially homogeneous but
temporally  oscillating reaction. Here we will focus on
the Hopf bifurcation, which is preferred for similar
diffusivities of the reacting substances, and on the
effects of a spatial modulated illumination
in one spatial dimension.
For this purpose,  the model for the two
dimensionless concentrations  $u(x,t)$ and $v(x,t)$ is
extended by a term describing a spatially varying
illumination $\phi(x)$, similar as in Ref.~\cite{Epstein:1999.1}:
\begin{subequations}
\label{Cfepstein}
\begin{eqnarray}
\partial_t u &=& a -cu -4 \frac{ uv}{1+u^2} - \phi + \partial_x^2 u \, , \\
\partial_t v &=& \sigma \left( cu - \frac{ uv}{1+u^2} + \phi + d \partial_x^2 v \right ) .
\end{eqnarray}
\end{subequations}
The constants $a, c, \sigma$ and $d$ denote
dimensionless parameters of the reaction diffusion model
and the  effect of an external illumination is introduced
through the field $\phi(x)$,
\begin{eqnarray}
\label{eq:2}
 \phi(x) = \phi_0 + M(x) \,.
\end{eqnarray}
 which can be identified  as the control
parameter of the system that is composed of a  spatially
homogeneous contribution $\phi_0$  and 
a spatially varying  part $M(x)$. $M(x)$
breaks the translational
symmetry of the system, and we  assume for reasons of simplicity 
a spatially periodic
modulation as described by
\begin{equation}
\label{eq:3}
M(x) = 2 G \cos{(2kx)},
\end{equation}
with the modulation amplitude  $2G$ and the modulation 
wave number $2k$.

With the two vectors \begin{eqnarray}
{\mathbf u} :=
    \begin{pmatrix} u  \\    v\end{pmatrix}  ,
\qquad
{\mathbf V} :=
    \begin{pmatrix}
                                       a - \phi \\    \sigma \phi 
    \end{pmatrix}  \,,  
 \end{eqnarray}
the matrix 
\begin{eqnarray}
\label{CfoperLN}
{\mathcal L} &:=&
    \begin{pmatrix}
                  \partial_t + c -\partial_x^2 & 0 \\
                 -\sigma c & \partial_t - \sigma d \partial_x^2
    \end{pmatrix},  \nonumber \\
\end{eqnarray}
and the nonlinear vector 
\begin{eqnarray}
{\mathbf N} &:=& \frac{uv}{1+u^2} 
    \begin{pmatrix}
                   - 4 \\ -\sigma 
    \end{pmatrix} ,
\end{eqnarray} 
a compact formulation of the two
basic equations  (\ref{Cfepstein})
becomes possible,
\begin{eqnarray}
\label{Cfepstein2} 
{\mathcal L} \mathbf{u}  = \mathbf{N}(\mathbf{u})  + {\mathbf V}  \,, 
\end{eqnarray}
which is  especially useful for
the amplitude expansion as outlined in the Appendix.

\section{\label{sec:3}Basic states and their stability}

The stationary basic state of  
the Lengyel-Epstein model is determined 
for a uniform illumination in  Sec.~\ref{sec:3.1} 
and for a spatially periodic 
illumination  in Sec.~\ref{sec:3.2}.
In both cases we also investigate its  stability
with respect to a bifurcation to an oscillating chemical reaction.

\subsection{\label{sec:3.1}The spatially homogeneous case  $M(x)=0$}

In the case  of a
homogeneous illumination, i.e. for $G=0$,
the stationary solution of the 
Eqs.~(\ref{Cfepstein}) 
is given by
\begin{align}
\label{Cfepsteinstat}
  u_0 &= \frac{ a -5 \phi_0}{5c} \, , &
  v_0 &= \frac{a\left( 1+u_0^2 \right )}{5u_0} \, .
\end{align}
It becomes unstable with respect to small oscillating perturbations
for an illumination strength
$\phi_0$ below a critical value $\phi_{0c}$, 
which is determined by a linear stability analysis.

\begin{figure}
  \includegraphics [width=0.45\textwidth] {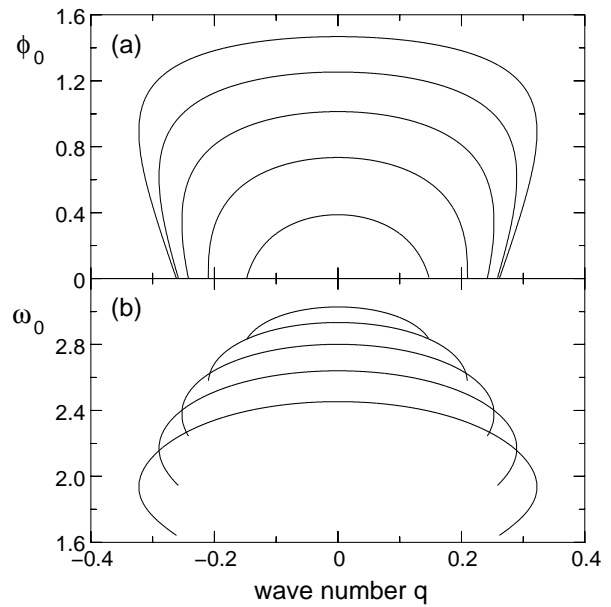}  
     \caption {
       The neutral curves $\phi_0(q)$  in part (a) correspond 
       starting from
       the top to increasing values  of $c=0.55, 0.65, 0.75, 0,85, 0.95$. 
       In part (b) the curves for the  Hopf frequency  $\omega_0(q)$
       along   $\phi_0(q)$ correspond starting 
       from the bottom to increasing values 
       of $c$.
        The remaining parameters are $a=12, \sigma=5, d=0.8$.
                  }
 \label{fig:1} 
\end{figure}

\begin{figure} 
  \includegraphics [width=0.45\textwidth] {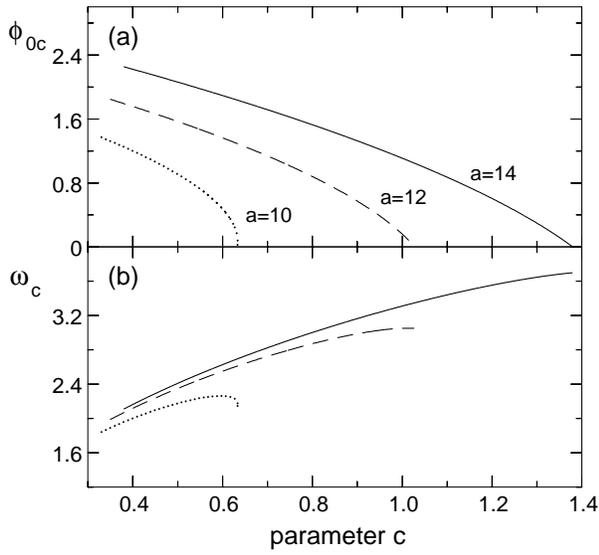} 
     \caption {
       In part (a)  the 
       threshold of the Hopf bifurcation, $\phi_{0c}=\phi_0(q=0)$, 
       and in part (b) the critical frequency
       $\omega_c=\omega_0(q=0)$ are  shown
       as a function  of the parameter $c$ and for three 
       different values of $a$.
       At small  values of $c$,  the Hopf bifurcation disappears and 
       the homogeneous state ${\bf {u}}_0$ becomes unstable with respect
       to a Turing instability. The data are determined for $\sigma=5$, $d=1$.
                  }
\label{fig:2}      
\end{figure}

For this analysis we start with
a  superposition of the  stationary, homogeneous
basic state $ \mathbf{u}_0$
and an infinitesimal perturbation $\mathbf{u}_1(x,t)$,
\begin{equation}
\mathbf{u} = \mathbf{u}_0+\mathbf{u}_1 =     \begin{pmatrix} u_0 \\ v_0  \end{pmatrix}  
                                                         +     \begin{pmatrix} u_1(x,t) \\ v_1(x,t) \end{pmatrix} \, ,
\end{equation}
as a solution of Eq.~(\ref{Cfepstein2}),  which is then linearized 
with respect to  $u_1$ and $v_1$. 
The resulting two coupled differential equations
\begin{eqnarray}
\label{Cflinstabu1}
    {\mathcal L} \mathbf{u}_1 = {\mathcal M}_0 \mathbf{u}_1 
\end{eqnarray}
have the constant coefficient matrix 
\begin{eqnarray}
{\mathcal M}_0 = 
    \begin{pmatrix}
                    -4 C_1 & -4 C_2 \\ -\sigma C_1 & -\sigma C_2  
    \end{pmatrix} 
 \end{eqnarray}
with the matrix elements
\begin{align}
C_1 &= \frac{v_0\left(1-u_0^2\right)}
                              {\left( 1+u_0^2 \right)^2} \, ,&   
C_2 &= \frac{u_0}{1+u_0^2}  \,\,. 
\end{align}
Equation~(\ref{Cflinstabu1}) may be further reduced by 
a mode ansatz of the form
\begin{equation}
\label{Cfstabu1v1}
 \mathbf{u}_1 =  A \begin{pmatrix} 1 \\ E_0  \end{pmatrix} 
                         \,\, e^{\lambda t + iqx} + {\rm c.c.}\,, 
\end{equation}
where  c.c. denotes the complex conjugate and 
$E_0$ describes the ratio between the amplitudes of the
two perturbations $u_1$ and $v_1$. The resulting 
two coupled linear equations
have only solutions for   a non-vanishing amplitude $A\not =0$,
if the  solubility  condition
\begin{eqnarray}
\label{CfdeterA}
{\rm det}
    \begin{pmatrix}
                  \lambda + c +q^2 + 4C_1 & 4C_2 \\
                 -\sigma c +\sigma C_1& \lambda + \sigma d \, q^2 +\sigma C_2
    \end{pmatrix} = 0\,
\end{eqnarray}
is fulfilled.  This condition determines
the two eigenvalues $\lambda(q)$ 
as functions of the wave number $q$ 
\begin{eqnarray}
\label{Cfeigenvalues}
\lambda_{\pm} &=&  \pm \frac{1}{2} \sqrt{  \left[ q^2(\sigma d+1) + 4C_1 +c+\sigma C_2 \right ]^2 -4 h(q^2) }
                                           \nonumber \\
                       && -\frac{1}{2} \left[ q^2(\sigma d+1) + 4C_1 +c+\sigma C_2 \right ] \, , 
\end{eqnarray}
with
\begin{eqnarray}
                h(q^2) = \sigma d q^4 +q^2 \left[ \sigma d(c+4C_1)+\sigma C_2 \right ] +
                       5 \sigma c C_2 \, \nonumber.
\end{eqnarray}
For a positive growth rate  ${\rm Re}(\lambda)$ and a finite imaginary part 
${\rm Im}(\lambda)\not = 0$,  the basic state $\mathbf{u}_0$ 
becomes unstable with respect to a Hopf bifurcation. 
The {\it neutral curve} of the wave-number dependent illumination strength
$\phi_0(q)$, which separates the stable from the unstable parameter range,
is determined by the  {\it neutral stability condition}
${\rm Re}[\lambda(q)]=0$.  It is  shown together with the
corresponding Hopf frequency  $\omega_0(q)$ 
for different values of the parameters $c$
in  Fig.~\ref{fig:1}.  Since a strong
illumination of the chemical reaction
suppresses the instability, the  homogeneous basic state $\mathbf{u}_0$ 
is unstable for a given value of $c$ within  the
area enclosed by the respective line in part (a) of  Fig.~\ref{fig:1}.
The maximum of each  neutral curve $\phi_0(q)$ is given
at $q=0$ that  determines
the critical illumination strength $\phi_{0c}$,  below
which the chemical reaction  becomes oscillatory.  
In this case  
the eigenvalues in Eq.~(\ref{Cfeigenvalues}) may be 
further simplified to  
\begin{eqnarray}
\label{Cfroots}
 \lambda_{\pm} &=& -\frac{1}{2} \left[ 4C_1 +c+\sigma C_2 \right ]   \nonumber \\
     && \pm \frac{1}{2} \sqrt{  \left[ 4C_1 +c+\sigma C_2 \right ]^2 -20 \sigma c C_2 } \, .
\end{eqnarray}
Since the parameters $c, C_2$ and $\sigma$ are all positive, also  
the product $\sigma c C_2$ 
is always positive and, therefore, 
the eigenvalues given by Eq.~(\ref{Cfroots}) are   either real 
with the same sign or
complex conjugate. The latter case occurs if  the condition
\begin{equation}
\label{Cfstabcond2}
\left[ c+ \frac{4v_0(1-u_0^2)}{(1+u_0^2)^2}+\frac{\sigma u_0}{1+u_0^2} \right ]^2 
-\frac{20\sigma c u_0} {1+u_0^2} < 0 
\end{equation}
is fulfilled and 
the stability of the ground state ${\mathbf u}_0$ is then determined by 
$\mbox{Re}(\lambda_{\pm})= \tau(a, c, \sigma, \phi_0)=-(4C_1 +c+\sigma C_2)/2$. 
The neutral  stability condition $\tau(a, c, \sigma, \phi_0)=0$ for the 
Hopf bifurcation is then in its 
explicit form given by
\begin{eqnarray}
\label{Cfphi0crit}
0 &=& 125 \phi_{0c}^3 + 25( a - 5\sigma) \phi_{0c}^2 + 25( 5c^2-a^2+2\sigma a) \phi_{0c} \nonumber \\
  && + a(3a^2 - 5\sigma a - 125c^2) \, , 
\end{eqnarray}
from which the 
critical illumination 
$\phi_{0c}(a, c, \sigma)$ may be determined.
The  Hopf frequency at this critical value is
described by the expression
\begin{eqnarray}
\label{Cfomegac}
 \omega_c = \pm \sqrt{5\sigma c C_2} = \pm \sqrt{ \frac{\sigma(a-5\phi_{0c})} 
                { 1 + \left( \frac{a-5\phi_{0c}}{5c} \right )^2 } } \,\, .
\end{eqnarray}
%
%
%
%
%
Both, the critical illumination $\phi_{ 0c}$
and the Hopf frequency $\omega_c$,  
are plotted in Fig.~\ref{fig:2} as functions of the parameter $c$ and for
three different  values of $a$. 
With increasing values of $c$,  the critical 
illumination decreases continuously up to the point
$\phi_{0c}=0$.  
For $c>c(\phi_{0c}=0)$
the stationary  and homogeneous
chemical reaction described by $\mathbf {u}_0$ is  always stable with
respect to small perturbations.
At small values of $c$, 
the Hopf bifurcation disappears and 
the basic state becomes unstable with
respect to a Turing instability.    
%
%
%
%

\subsection{\label{sec:3.2}Basic state in the presence of $M(x)$}

In order to determine in case of a spatially 
periodic illumination 
$\phi(x)=\phi_0+M(x)$ 
the stationary basic state $\hat{{\bf {u}}}_0=(\hat u_0,\hat v_0)$ 
of Eq.~(\ref{Cfepstein2}),   
we use its   time-independent part 
in the following form  
\begin{subequations}
\label{FostatLEmod}
\begin{align}
  0&= a -5cu - 5\phi(x) +\partial_x^2 u -4d\partial_x^2 v, \\
 0&= a(1+u^2)  - 5 uv + (1+u^2)\left( \partial_x^2 u + d\partial_x^2 v \right ) .
\end{align}
\end{subequations}
These inhomogeneously  forced differential equations
may be solved by the following Fourier ansatz 
for the two fields $\hat u_0(x)$ and $\hat v_0(x)$:
\begin{align}
\label{Fobsuv}
  \hat u_0 &= \sum_{l=-M}^M U_l \, e^{il2kx} , & 
\hat v_0  &= \sum_{l=-M}^M V_l \, e^{il2kx} 
\end{align}
with an appropriate number $M$ and the Fourier amplitudes
$U_l$ and   $V_l$ of the expansion. 
The magnitude of these amplitudes for $l\not =0$ 
is essentially
determined by the forcing amplitude $G$.
Since $\hat u_0$ and $\hat v_0$
are real functions, 
we assume real amplitudes with $U_l=U_{-l}$ and
$V_l=V_{-l}$, respectively.
Substituting the ansatz~(\ref{Fobsuv})  into Eqs.~(\ref{FostatLEmod}) 
and, after projecting the equations onto 
$\int dx \, \exp{(-ij2kx)}$
in order to eliminate the $x$-dependence, we obtain a set of
coupled algebraic equations for the determination of the 
unknown Fourier amplitudes:
\begin{subequations}
\label{FoLEfoueq}
\begin{eqnarray}
\label{FoLEfoueqa}
 0 &=& \left( a - 5 \phi_0 \right) \delta_{j,0} -5c U_j 
-5G \left( \delta_{j,1}+\delta_{j,-1} \right) \nonumber \\
            && - (2jk)^2 U_j +4d(2jk)^2 V_j \, , \\
\label{FoLEfoueqb}
0&=&  a \delta_{j,0} + a \sum_l U_l U_{j-l} -5 \sum_l U_l V_{j-l} 
                                                                            -(2jk)^2 U_j   \nonumber \\ 
          && - d(2jk)^2 V_j - \sum_l \sum_m U_l U_m U_{j-l-m} (2mk)^2 \nonumber \\
 && - d\sum_l \sum_m U_l U_{j-l-m} V_m (2mk)^2 
 \end{eqnarray}
\end{subequations}
with $j=-M \ldots M$.
All sums in Eq.~(\ref{FoLEfoueqb}) run  from $-M$ to $M$ and
the  system of nonlinear equations in the amplitudes 
$U_j$ and $V_j$ 
can be solved by standard numerical methods. The
basic spatially dependent 
solutions are then evaluated via  Eq.~(\ref{Fobsuv}). One should note  that
according to Eq.~(\ref{FoLEfoueqa}),  
the Fourier amplitude $U_0=(a-5\phi_0)/(5c)$ 
corresponding to the spatially homogeneous contribution to  $\hat u_0(x)$
is not changed by the forcing, cf.~Eq.~(\ref{Cfepsteinstat}). 
%
%
\begin{figure} 
  \includegraphics [width=0.47\textwidth] {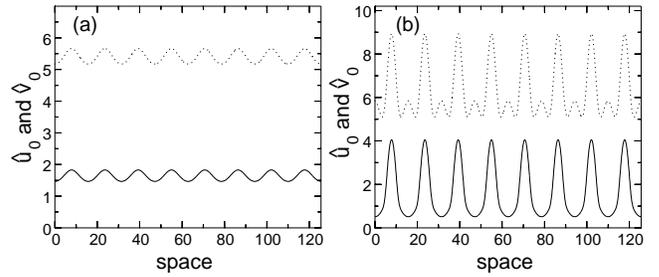} 
     \caption {
       The basic solutions $\hat u_0$ (solid line) and $\hat v_0$ (dotted  line)
        of the Lengyel-Epstein model for a spatially modulated illumination
       rate $\phi(x)=\phi_0 +2G\cos{(2kx)}$ with wave number $k=0.2$
       are shown in part (a) for 
       $G=0.04$ and in part (b) for $G=0.4$.
        Parameters are $a=12, c=0.55, d=0.8, \phi_0=1.5$. 
                  }
\label{fig:3}
\end{figure}  
%
%

For a given set of parameters,  
the two solutions $\hat u_0$ and $\hat v_0$
as given by Eq.~(\ref{Fobsuv}) are plotted 
in Fig.~\ref{fig:3}(a) for the modulation amplitude 
$G=0.04$  and in part (b) for $G=0.4$ while  the modulation
wave  number is given by $k=0.2$.
The field $\hat u_0(x)$ is pictured as a solid line 
and $\hat v_0(x)$ as a dotted line. The spatial profile of 
the solutions shown in part (a) is dominated by  
the wave number of the forcing $M(x) \propto \cos{(2kx)}$. 
For increasing forcing amplitudes $G$,   the weights of the
higher harmonic amplitudes in the expansions
given in Eq.~(\ref{Fobsuv})
are  amplified and the basic state $\hat {\mathbf {u}}_0$
becomes  fairly anharmonic as illustrated in part (b).
Note the different scales in part (a) and (b).

\subsection{\label{sec:3.3}Threshold of the Hopf bifurcation  in the presence of $M(x)$}

The spatially periodic basic state of the Lengyel-Epstein model
as described by $\hat {\mathbf {u}}_0=(\hat u_0(x),\hat v_0(x))$ 
becomes unstable against
infinitesimal perturbations $\mathbf {w}=(w_1,w_2)$
below a critical illumination rate $\phi_0< \phi_{0c}(G,k)$.  
In order to determine this critical  value, the basic state is
separated from the small perturbation by the ansatz
\begin{eqnarray}
\label{FOLEuv}
  \mathbf {u} = \hat {\mathbf {u}}_0 + \mathbf {w}\, .
\end{eqnarray}
After linearizing the basic equation~(\ref{Cfepstein2}) 
with respect to $\mathbf {w}$,
one obtains the following equation of motion
\begin{eqnarray}
\label{FoLElinstabeq}
 {\mathcal L} \mathbf{w} = \hat {{\mathcal M}}_0 \mathbf{w} ,
\end{eqnarray}
with the coefficient matrix
\begin{eqnarray}
\hat{{\mathcal M}}_0 = 
    \begin{pmatrix}
                    -4 \hat C_1 & -4 \hat C_2 \\ 
                    -\sigma \hat C_1 & -\sigma \hat C_2  
    \end{pmatrix} 
 \end{eqnarray}
and the abbreviations
\begin{align}
\label{FoC1C2x}
  \hat C_1(x) &= \frac{ \hat v_0(1- \hat u_0^2)}{(1+ \hat u_0^2)^2} \,, 
  & \hat C_2(x) &= \frac{\hat u_0}{ 1+ \hat u_0^2} \, .
\end{align}
Equation~(\ref{FoLElinstabeq}) has formally the same form
as for the homogeneous case given by Eq.~(\ref{Cflinstabu1}). 
Since Eq.~(\ref{FoLElinstabeq}) has  spatially varying coefficients 
with a periodicity given by the forcing wave number, $2k$, the following
Floquet-type ansatz for the small perturbations may be chosen with
a complex parameter  $\lambda$:
\begin{eqnarray}
\label{FoLEw1w2}
\mathbf{w} = e^{\lambda t} \sum_{l=-N}^{N}   \begin{pmatrix}  F_{l}       \\  H_{l} \end{pmatrix} 
                                     e^{ilkx} + {\rm c.c.}   \,\quad .
\end{eqnarray}
Substituting the ansatz~(\ref{FoLEw1w2}) into Eq.~(\ref{FoLElinstabeq}) 
and using additionally the Fourier representation  of 
$\hat C_1(x)$ and $\hat C_2(x)$,
\begin{align}
\hat C_1&=\sum_{l=-M}^M C_l^{(1)} e^{2ilkx}\, , & \hat C_2&= \sum_{l=-M}^M C_l^{(2)} e^{2ilkx} \, ,
\end{align}
all terms can be sorted with respect to the linearly independent
 exponential functions. 
To transfer the linear equation~(\ref{FoLElinstabeq})
into an eigenvalue problem
for the constant coefficients $F_l$ and $H_l$, one
has to eliminate the remaining dependence on $x$ by projecting
the equations onto $\int dx\,e^{-ijkx}$.
One finally ends up with the following system of equations:
\begin{subequations}
\label{FoLEFjHj}
\begin{eqnarray}
\lambda F_j &=& -cF_j -4 \sum_{l=-M}^M C_l^{(1)} F_{j-2l} -4 \sum_{l=-M}^M C_l^{(2)} H_{j-2l}
                                 \nonumber \\ 
                             && -(jk)^2 F_j  \, , \\
\lambda H_j &=& \sigma \left[ -\sum_{l=-M}^M C_l^{(2)} H_{j-2l}   -d(jk)^2 H_j + cF_j  \right.\nonumber \\
                            &&       \left.  - \sum_{l=-M}^M C_l^{(1)} F_{j-2l} \right ] 
\end{eqnarray}
\end{subequations}
with $j=-N \ldots N$. These equations can be written in a more compact form
as two coupled sets of equations 
\begin{subequations}
\label{FoLEeigFH}
\begin{eqnarray}
  \lambda \mathbf{F} &=& {\mathcal A_1} \mathbf{F} + {\mathcal I_1} \mathbf{H} \, ,\\
 \lambda \mathbf{H} &=& {\mathcal A_2} \mathbf{H} + {\mathcal I_2} \mathbf{F} \, ,
\end{eqnarray}
\end{subequations}
where $ {\mathcal A_i}$ and  ${\mathcal I_i}$  $(i=1,2)$ denote matrices
of the dimension $(2N+1) \times (2N+1)$
and $\mathbf{F}$ and $\mathbf{H}$ include the $(2N+1)$ Fourier amplitudes
of $w_1$ and $w_2$, respectively.
Additionally, Eqs.~(\ref{FoLEeigFH}) can be formally rewritten as
an eigenvalue problem
\begin{eqnarray}
\label{FoLEeigvalue}
             {\mathcal C}  \mathbf{\Psi}  =  \lambda  \mathbf{\Psi} 
\end{eqnarray}
with 
\begin{eqnarray}
 \mathbf{\Psi} := 
    \begin{pmatrix}
                   \mathbf{F} \\ \mathbf{H}
    \end{pmatrix}  \, \quad {\rm and} \quad
 {\mathcal C} :=
    \begin{pmatrix}
                  {\mathcal A_1}  & {\mathcal I_1}  \\
                  {\mathcal I_2} &  {\mathcal A_2} 
    \end{pmatrix} \, .
\end{eqnarray}
The matrix ${\mathcal C}$ has the dimension $(4N+2) \times (4N+2)$.
From Eqs.~(\ref{FoLEFjHj}) one recognizes that 
the even and odd indices $j$ are actually decoupled giving rise to two
independent thresholds. These are the harmonic threshold $\phi_{0c}^h$
corresponding to harmonic perturbations $w_i^h(x+\pi/k)=w_i^h(x)$ ($i=1,2$)
with respect to the forcing $M(x)$ 
and the subharmonic threshold $\phi_{0c}^{sh}$ corresponding
to subharmonic perturbations $w_i^{sh}(x+\pi/k)=-w_i^{sh}(x)$.
The larger one of these two thresholds determines whether spatially
harmonic or subharmonic patterns emerge from
the basic state $\hat {\mathbf {u}}_0$  via a Hopf bifurcation.

Technically we solve the threshold problem as follows:
(i) We keep besides the illumination $\phi_0$  all the other parameters
$a, c, d, \sigma, G, k$ fixed and determine the basic states $\hat u_0$ and
$\hat v_0$  by solving Eqs.~(\ref{FoLEfoueq}). \\
(ii) Since the 
functions  $\hat C_1(x)$ and $\hat  C_2(x)$ depend
according to  Eq.~(\ref{FoC1C2x}) nonlinearly
on the basic state $\hat {\mathbf {u}}_0$, 
they are numerically evaluated 
on a discrete lattice and the
Fourier amplitudes $C_l^{(1)}$ and $C_l^{(2)}$, as required 
in Eq.~(\ref{FoLEeigvalue}), are 
obtained by a numerical Fourier transformation.    \\
(iii) The eigenvalue spectrum of the matrix $\mathcal C$ determines 
whether the basic state $\hat {\mathbf {u}}_0$  is  stable, 
i.e.  if ${\rm Re}(\lambda)<0$  for all eigenvalues,
or unstable,  i.e.  if ${\rm Re}(\lambda)>0$ at least for one
eigenvalue $\bar \lambda$.
Here $\bar \lambda$ denotes the eigenvalue with the largest growth rate.
The illumination $\phi_0$ is varied until the {\em neutral stability} condition
${\rm Re}(\bar \lambda)=0$ is satisfied and this specific
value of $\phi_0$ determines the  critical illumination $\phi_{0c}$,  
while ${\rm Im}(\bar \lambda)=\pm \omega_c$  gives
the Hopf frequency at threshold.

\begin{figure} 
\includegraphics [width=0.45\textwidth] {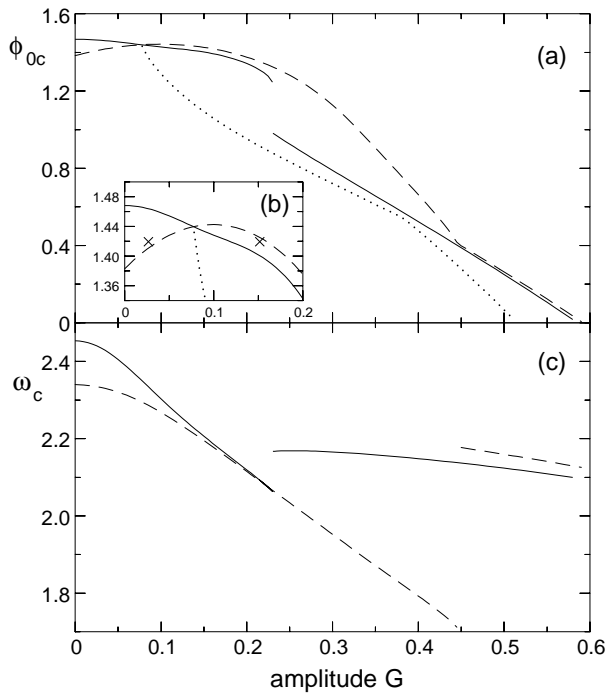} 
     \caption {
       In part (a) the threshold  $\phi_{0c}^{h}$ of   the  harmonic instability  
       (solid line) and the threshold $\phi_{0c}^{sh}$ of the subharmonic one
       (dashed line) are shown as functions
       of the forcing amplitude $G$.  
       Spatially subharmonic patterns are preferred in a wide range of $G$
       with $\phi_{0c}^{sh} > \phi_{0c}^{h}$.
       The dotted line indicates the lower end of the 
       existence region of
       subharmonic patterns.
       The inset (b) displays the behavior of the thresholds 
       for small values of $G$
       in order to enlarge the crossing region of the two curves  $\phi_{0c}^{h}(G)$ 
       and $\phi_{0c}^{sh}(G)$. 
       The two crosses in the inset mark the points at 
       which some results of numerical 
       simulations of Eqs.~(\ref{Cfepstein}) will be  presented (see Fig.~\ref{fig:6}). 
       The Hopf frequency along the two threshold curves is  shown in part (c).
       Parameters are $a=12,~ c=0.55,~ \sigma=5,~ d=0.8,~ k=0.2$.
                  }
\label{fig:4}
\end{figure}

\begin{figure} 
\includegraphics [width=0.45\textwidth] {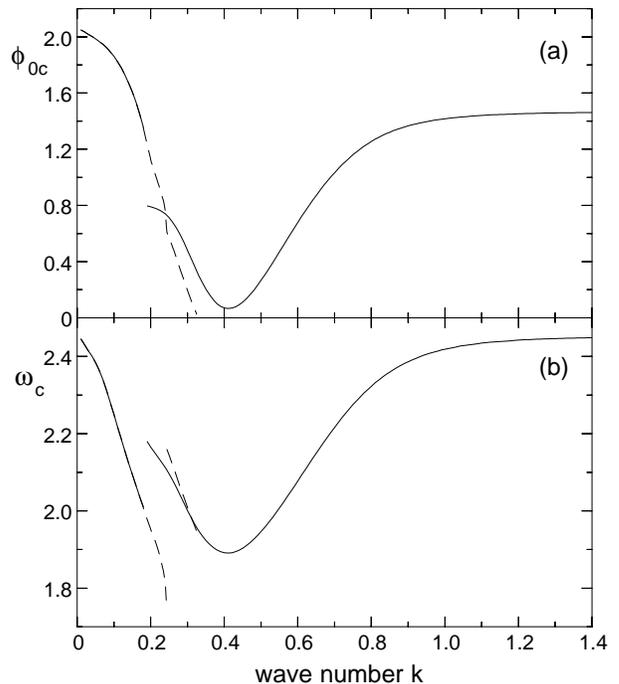} 
     \caption {
       The thresholds for the harmonic instability $\phi_{0c}^{h}$ (solid line) 
       and the subharmonic one 
       $\phi_{0c}^{sh}$ (dashed line) are shown as functions of the forcing wave number $k$
       for a modulation strength $G=0.3$. In a small range of $k$ the subharmonic
       solution is preferred beyond the instability, 
       $\phi_{0c}^{sh} > \phi_{0c}^{h}$.
       In the limit $k\to \infty$
       the harmonic threshold converges to $\phi_{0c}(G=0)=1.47$.
       In the opposite limit,  i.e.  $k\to 0$,
       both thresholds coincide  
       and the critical illumination is given by $\phi_{0c}(G=0) +2G$.
       The Hopf frequency is shown in part (b).
       The remaining parameters 
       are the same as used in  Fig.~\ref{fig:4}.
                  }
\label{fig:5}
\end{figure}

Some results for the harmonic and subharmonic
threshold as well as for the corresponding 
Hopf frequency are presented  in Fig.~\ref{fig:4}
as functions of the amplitude $G$ and in Fig.~\ref{fig:5} 
as functions of the wave number $k$. 
Here, $\phi_{0c}^h,~ \omega_c^h$ are pictured as solid lines and
$\phi_{0c}^{sh}, ~\omega_c^{sh}$ as dashed lines.
The harmonic solutions have according to Fig.~\ref{fig:4}(a) 
the higher  threshold for small forcing amplitudes $G$ 
and from the  basic state,  a Hopf bifurcation occurs
which is spatially
harmonic with respect to the external modulation.  
However, the harmonic threshold drops below the subharmonic one
for all values $G>0.077$  and the bifurcation 
from the basic state  is changed to a spatially
subharmonic pattern. 
%
%
%
%
The upper envelope of both threshold curves is   the instability border
below which the basic state $\hat {\mathbf {u}}_0$
becomes  unstable against small oscillating perturbations.
Subharmonic patterns are expected to occur 
in numerical simulations with  random initial  solutions
within the area   enclosed by the subharmonic threshold
and by the dotted line in part (a) of the figure.
Part (b) displays the two thresholds in the range
of small values of $G$ in order to magnify the neighborhood of
the  intersection between $\phi_{0c}^h(G)$ and $\phi_{0c}^{sh}(G)$.
The two crosses in part (b) indicate the parameter values of $\phi_0$ and $G$,  for which
some nonlinear solutions  of the Lengyel-Epstein model~(\ref{Cfepstein})
will be  presented in  Fig.~\ref{fig:6}.
The Hopf frequency along the two thresholds in part (a) is depicted in part (c).
%
%
%
%
As can be seen from Fig.~\ref{fig:5}(a), 
the thresholds for the harmonic and 
subharmonic solution differ only slightly for small forcing wave numbers $k$ 
and in the limiting case $k \to 0$, where 
the critical illumination is given by 
$\phi_{0c}=\phi_{0c}(G=0)+2G$, both thresholds coincide.
On the other hand, for large forcing wave numbers where  the modulation wavelength
becomes smaller than the diffusion length, 
the system averages over the fast spatial oscillations $\sim 1/k$,   and the threshold
approaches its unmodulated value  $\phi_{0c}(G=0)$ from below.
The harmonic threshold $\phi_{0c}^{h}(k)$ has a pronounced minimum
at  $k\approx 0.4$. Here 
the bifurcation is almost suppressed, i.e.
only a very weak  illumination forces an instability
of the stationary state $\hat {\mathbf {u}}_0$.
For long-wavelength modulations $ k \ll 1$,  the critical
illumination is given by $\phi_{0c}>\phi_{0c}(G=0)$ and, 
therefore,  the Hopf bifurcation already occurs in a  range
of the illumination $\phi_0$, where it
does not appear  without modulation.
Again, in the $k$-range where the harmonic threshold drops below the
subharmonic one, the stationary basic state becomes  unstable against
spatially subharmonic perturbations.
The corresponding Hopf frequency is  displayed in part (b).

\section{\label{sec:4}Nonlinear solutions}

The determination of the time evolution of the nonlinear solutions of
Eqs.~(\ref{Cfepstein}) for spatially periodic boundary conditions
is performed  by a pseudospectral method.
From a numerical point of view, 
it has been proven useful to separate the stationary  basic state $\hat {\mathbf {u}}_0$  
from the oscillatory contribution $\mathbf{w}$
in order to simulate the equations of motion.
After inserting the ansatz~(\ref{FOLEuv}) in the Lengyel-Epstein model~(\ref{Cfepstein}),
one obtains the following governing equations for the two fields $w_1$ and $w_2$:
\begin{subequations}
\label{Fosimw1w2}
\begin{eqnarray}
 \partial_t w_1 &=& -cw_1 + \partial_x^2 w_1 \nonumber \\
                 &&  -4 \left[ \frac{(\hat u_0+w_1)(\hat v_0+w_2)}{1+(\hat u_0+w_1)^2} 
                             - \frac{\hat u_0 \hat v_0}{1+\hat u_0^2} \right ]  , \\
 \partial_t w_2 &=& \sigma (cw_1 + d\partial_x^2 w_2) \nonumber \\
                                 &&  -\sigma \left[ \frac{(\hat u_0+w_1)(\hat v_0+w_2)}{1+(\hat u_0+w_1)^2} 
                             - \frac{\hat u_0 \hat v_0}{1+\hat u_0^2} \right ]  . 
\end{eqnarray}
\end{subequations}
The effect of the modulation $M(x)$ enters these equations
via the basic states  $\hat u_0$ and $\hat v_0$. 
A great deal of simulations  with random initial conditions
were performed in order to verify the onset of the Hopf bifurcation numerically.
The results are in excellent agreement 
with the threshold  curves 
for $\phi_{0c}^h(G)$ and $\phi_{0c}^{sh}(G)$, respectively, 
in Fig.~\ref{fig:4}(a).

\begin{figure} 
 \includegraphics [width=4.2cm] {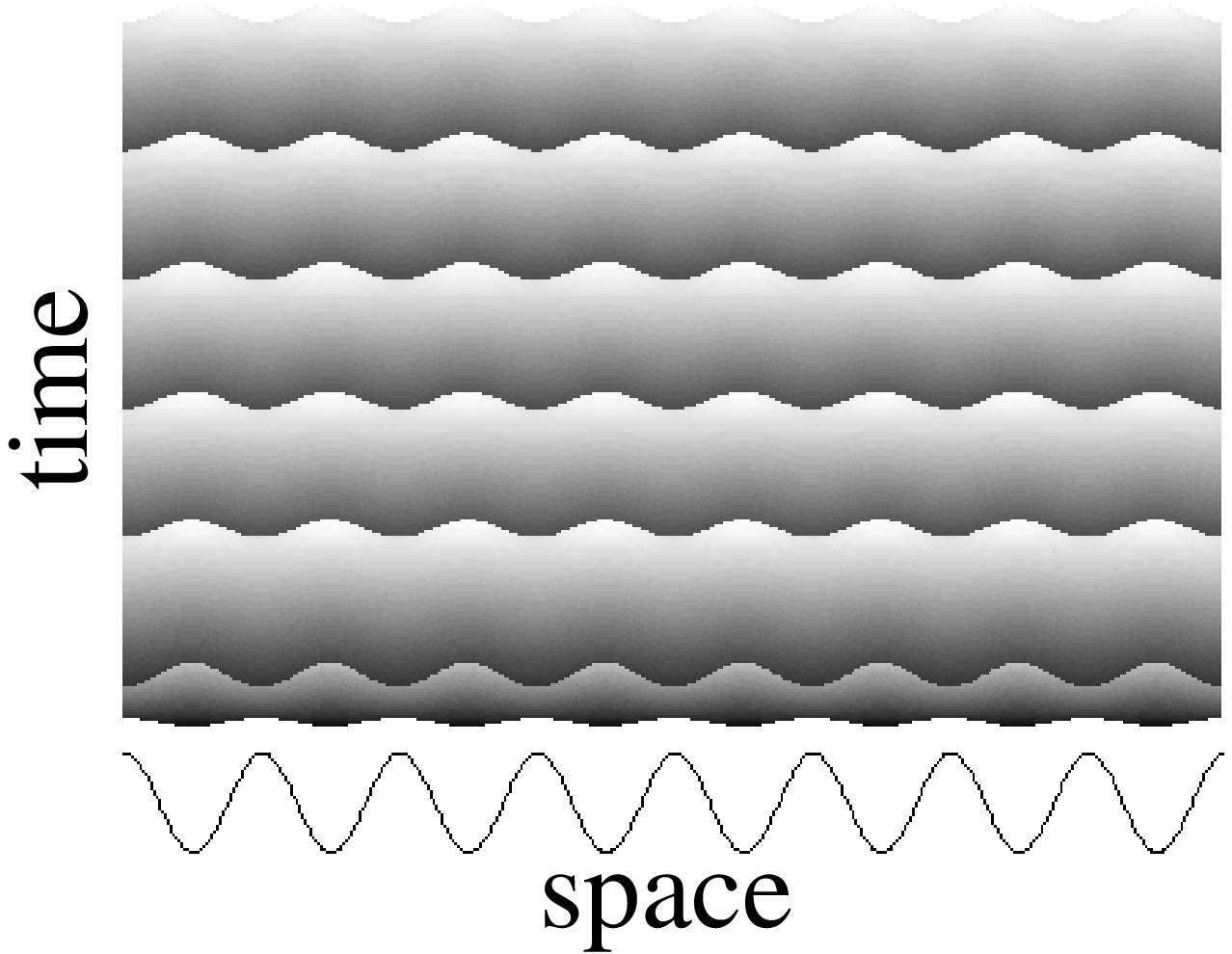}       
 \includegraphics [width=4.2cm] {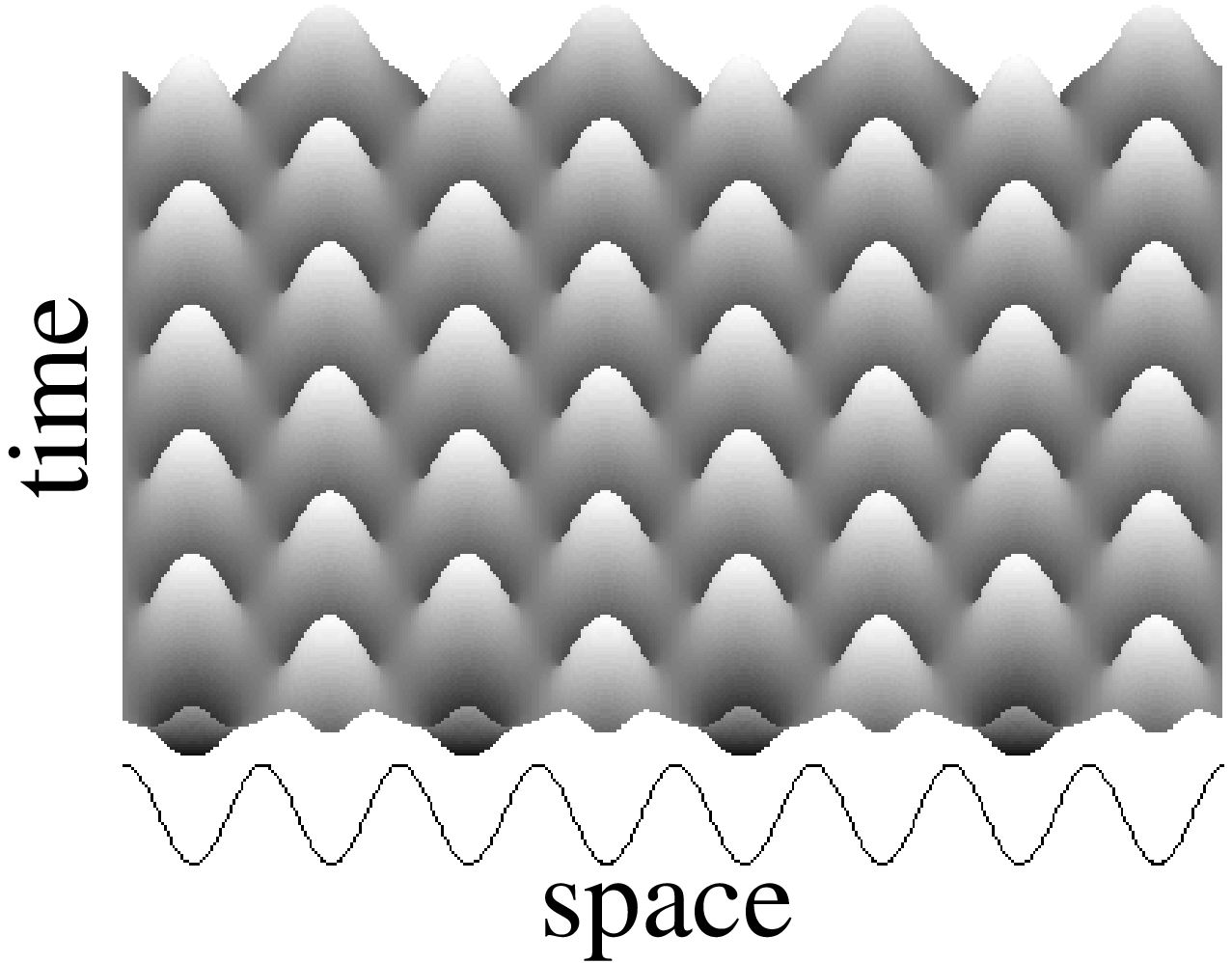}   \\
 \includegraphics [width=4.2cm] {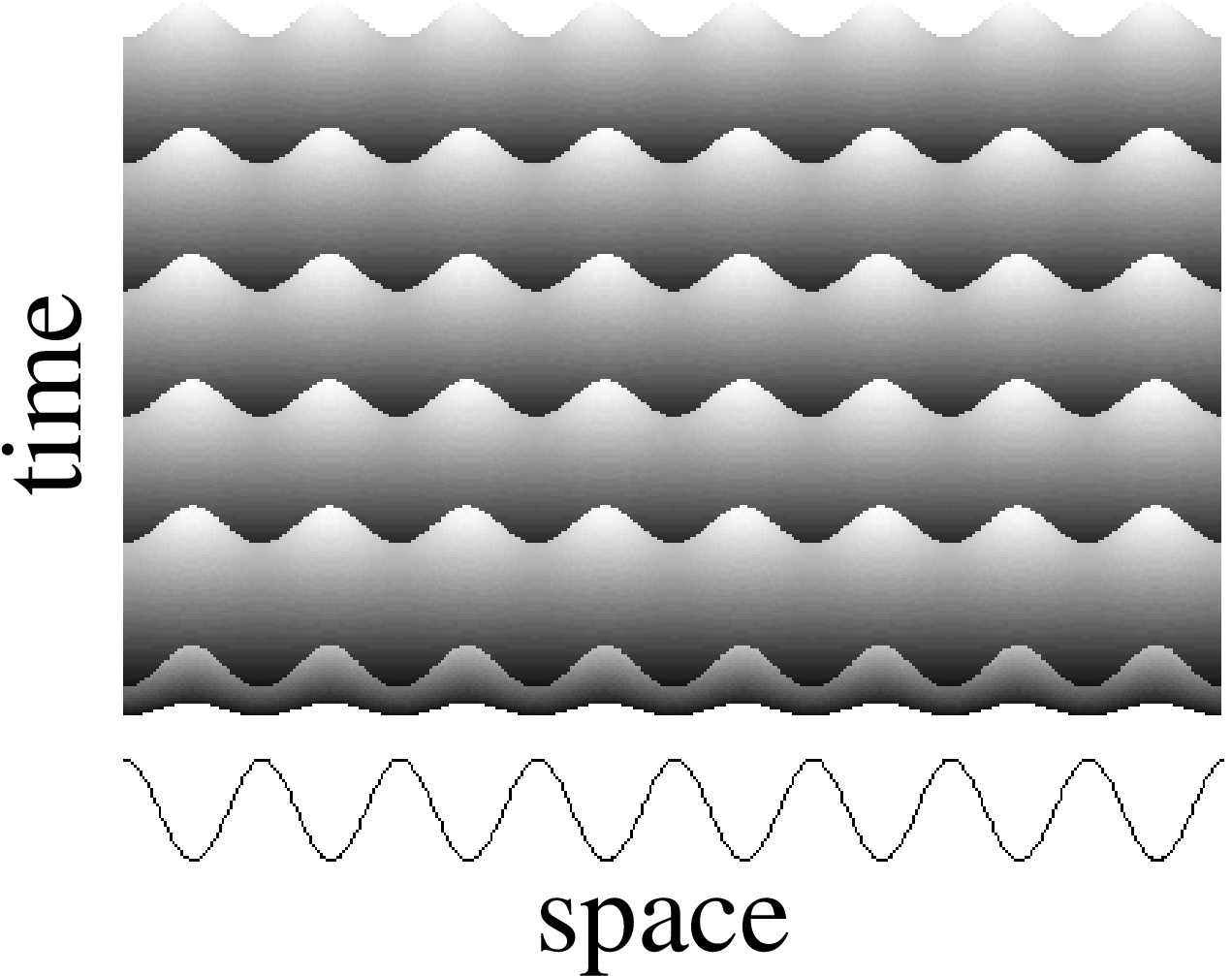}    
 \includegraphics [width=4.2cm] {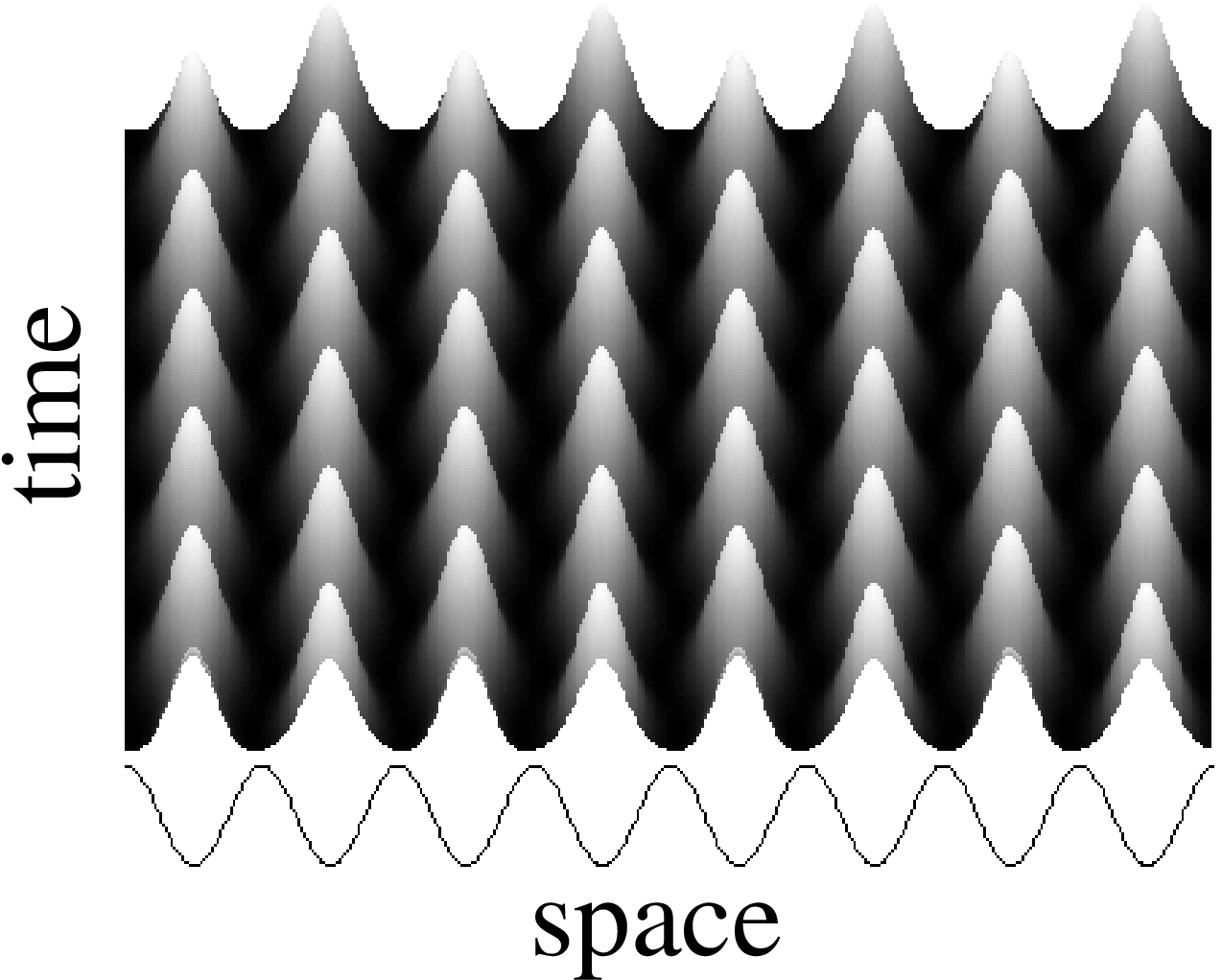}                                                                                                      
     \caption {                                                                                            
       The left column shows
       harmonic solutions for the fields
        $w_2^h$ (top) and $v=\hat v_0+w_2^h$ (bottom) 
       and the right column subharmonic solutions for the same fields,  i.e.
       $w_2^{sh}$ (top) and $v=\hat v_0+w_2^{sh}$ (bottom).
        Solid lines indicate  the spatial modulation $M(x)$.
       The illumination is given by $\phi_0=1.42$ and the forcing amplitude by $G=0.025$
       in part (a) and by $G=0.15$ in part (b).
        Further parameters are $a=12, c=0.55, \sigma=5, d=0.8, k=0.2$.
        The parameter values  in the $\phi_0-G$ plane 
        are marked by crosses in Fig.~\ref{fig:4}(b).
                  }                                                                                          
\label{fig:6}
\end{figure}

Two types of nonlinear solutions obtained 
by numerical simulations
are presented in Fig.~\ref{fig:6}, where 
the left column  shows the time evolution 
of a  spatially harmonic solution 
for the field $w_2^h$ (top) as well as  for the superposition
$v=\hat v_0+w_2^h$ (bottom) 
occurring at $\phi_0=1.42$ and $G=0.025$.
The same fields are shown in the right column, 
i.e.  $w_2^{sh}$ (top) 
and $v=\hat v_0+w_2^{sh}$ (bottom),  
in the case of subharmonic solutions occurring at $G=0.15$.
The parameters for these simulations in the  $\phi_0-G$ plane
are marked by the crosses
in Fig.~\ref{fig:4}(b).
%
%
%
%
The solution $w_2^h(x,t)$  describes 
spatially modulated oscillations whose spatial profile 
becomes more pronounced when the basic solution $\hat v_0$
is included.
The time evolution of the field $w_2^{sh}$ 
resembles that of a standing wave with
twice the wavelength of the forcing and,  indeed,
it is well described by a superposition
of three standing waves 
$w_2^{sh}(x,t) = \sum_{j=1}^{3} B_j \sin{(\omega_c^{sh} t+\varphi_j)}\sin{[(2j-1)kx]}$ 
with real amplitudes $B_j$ and phases $\varphi_j$.
On the bottom right the 
full solution $v=\hat v_0+w_2^{sh}$ of the Lengyel-Epstein model is shown. 
Note, the basic solutions $\hat v_0,~ \hat u_0$ have the same periodicity
as the spatial modulation.
The nonlinear solutions for the  other fields $w_1$ and $u=\hat u_0+w_1$
look very much like the
ones shown in Fig.~\ref{fig:6} and they are therefore
not presented here.
Starting the simulations with  spatially subharmonic states 
and decreasing the illumination $\phi_0$ continuously,
the subharmonic pattern  is stable against small perturbations
up to  the dotted line in Fig.~\ref{fig:4}(a).
On exceeding this border line,  
the subharmonic solution  becomes  unstable
and spatially harmonic patterns emerge.

Close to the threshold of the Hopf bifurcation, where
the amplitude of the oscillations is small, 
the dynamics
of the Lengyel-Epstein model may be described  by a so-called
amplitude equation as discussed in the next section.

\section{\label{sec:5}Amplitude equation}

Below the critical value $ \phi_{0c}$ of the
control parameter $\phi_0$, 
the basic state
$\mathbf{u}_0=(u_0,v_0)$ of the Lengyel-Epstein 
model~(\ref{Cfepstein})
becomes linearly unstable 
against small oscillatory perturbations
$\mathbf{u}_1$ as 
described in
Sec.~\ref{sec:3.1}.
The magnitude of  $\mathbf{u}_1$ beyond threshold
is restricted by nonlinear terms in $\mathbf{u}_1$
which are of cubic order for  a supercritical bifurcation. In this
case one
may derive a universal amplitude equation for the slowly varying amplitude $A(x,t)$  
in order to describe the dynamics  close to the threshold \cite{CrossHo}.
The technique for this derivation is a multiple scale analysis where
the full solution  $\mathbf{u}_1$
is decomposed into a fast varying oscillation 
 $\propto \exp{(i\omega_c t)}$
with the frequency $\omega_c$ 
and a spatially and temporally slowly varying amplitude $A(x,t)$:
$\mathbf{u}_1=A(x,t)\mathbf{u}_e \exp{(i\omega_c t)} + {\rm c.c.}$.

The usage of  amplitude equations
is a well established method 
to characterize  the universal properties of a
pattern at small amplitudes close to its threshold. 
A particularly  well-known amplitude equation is found  for
a spatially homogeneous
Hopf bifurcation, which has been investigated very intensively 
over the recent decades and a recent
review of  this subject is given by Ref.~\cite{Aranson:02.1}.

\begin{figure} 
  \includegraphics [width=0.47\textwidth] {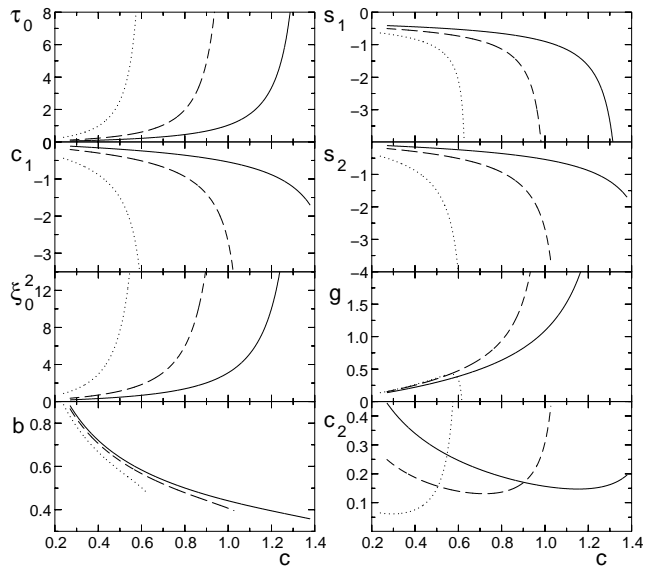} 
     \caption {                                                                                                 
       The coefficients of the amplitude equation~(\ref{CfMGLE})
       are plotted as functions of the parameter $c$ of the 
       Lengyel-Epstein model and for three different values $a$:
       $a=14$ (solid lines),  $a=12$ (dashed lines) and  $a=10$ 
       (dotted lines).
       The critical illumination $\phi_{0c}$ and the Hopf frequency $\omega_c$
       along these curves are shown in Fig.~\ref{fig:2}.
       Further parameters are $\sigma=5, d=1$.
                  }                                                                                             
\label{fig:7}
\end{figure}

For  the derivation of the amplitude equation one introduces
as an expansion parameter the small distance to the
threshold $\varepsilon=\frac{\phi_{0c}-\phi_0}{\phi_{0c}}$ and
the expansion holds in the range $A \propto \varepsilon^{1/2}$.
Here we assume additionally  that the modulation $M(x)$ is also of
the order $\varepsilon$. For this case  we generalize the amplitude
equation for  an oscillatory bifurcation 
by including the spatial modulation $M(x)$:
\begin{eqnarray}
\label{CfMGLE}
 \tau_0 \partial_t A &=& \varepsilon \left( 1+ic_1 \right) A+ \xi_0^2 \left( 1+ib \right) \partial_x^2 A
                              \\    
                      && + s_1 \left(1+i s_2 \right) M A 
                     -g\left(1+ic_2\right) \mid A \mid^2 A \, . \nonumber  
\end{eqnarray}
An explicit  derivation of this equation
from the basic Eqs.~(\ref{Cfepstein})
is given in the Appendix and a special case of it is  given
in Ref.~\cite{Zimmermann:96.2}.
All coefficients in Eq.~(\ref{CfMGLE}) describe physical quantities,
such as the relaxation time $\tau_0$, the linear and nonlinear frequency shift
$c_1$ and $c_2$, respectively, the coherence length $\xi_0$ and
the linear frequency dispersion $\xi_0^2b$.
For $g>0$ the bifurcation is supercritical and otherwise subcritical.
The analytical expressions for all these  coefficients 
in terms 
of the parameters of the basic equations~(\ref{Cfepstein})
are rather lengthy and, instead of giving their analytical forms,
we have plotted them in Fig.~\ref{fig:7}
as functions of the parameter $c$ and for three different values of $a$.

It is worthwhile mentioning  that
apart from the coefficients $s_1$ and $s_2$, all the other linear
coefficients can be calculated from the dispersion relation
$\lambda(\phi_0, q^2,\ldots)= {\rm Re}(\lambda) \pm i {\rm Im}(\lambda)$
of the Lengyel-Epstein model given in Eq.~(\ref{Cfeigenvalues}) 
by the following expressions \cite{CrossHo,Zimmermann:93.1}
\begin{subequations}
\begin{align}
\tau_0 &= \frac{1}{\phi_{0c} \partial {\rm Re}(\lambda)/\partial \phi_0}~, &
     c_1& = \phi_{0c}\tau_0\, \frac{\partial \omega} {\partial \phi_0}\,,  \\
  \xi_0^2 & =  \frac{1}{2 \phi_{0c}} \frac{\partial^2 \phi_0}{\partial q^2}~ , 
 & b&  = -\frac{\tau_0}{2 \xi_0^2}  \frac{\partial^2 \omega}{\partial q^2} ~.
\end{align}
\end{subequations}
Here $\omega={\rm Im}(\lambda)$ and  the
 derivatives are evaluated at the critical values
$\phi_{0c}, q_c, \omega_c$.
For a vanishing modulation these coefficients 
describe the linear properties
of the Lengyel-Epstein model near threshold.
It is however indispensable to carry out the  perturbation expansion
in order to  determine the linear coefficients $s_1$ and $s_2$ as well as 
the nonlinear coefficients $g$ and $c_2$
as functions  of the parameters of the basic equations.

The term $i\varepsilon c_1 A$ in Eq.~(\ref{CfMGLE})
can be removed by the transformation $\widetilde A = e^{-i\varepsilon c_1 t} A$. 
For convenience we can scale out the coefficients $\tau_0, \xi_0, s_1, g$ 
by a suitable choice of time, space  and amplitude scales
\begin{align}
t' &= t / \tau_0  \, ,&  x' &= x/ \xi_0  \, , \nonumber \\
A' &= g^{1/2} A \, ,& G'&= s_1 G \, ,
\end{align}
and one obtains the following rescaled amplitude equation:
\begin{eqnarray}
\label{Foampe}
  \partial_t A &=& \varepsilon A+ \left( 1+ib \right) \partial_x^2 A   + \left(1+is_2 \right) M A
                  \nonumber \\
              &&  -\left(1+ic_2\right) \mid A \mid^2 A \,, 
\end{eqnarray}
where we have kept for simplicity the same symbols for the scaled quantities.
The amplitude equation is invariant under an arbitrary phase
transformation as $A\rightarrow A \exp{(i\psi)}$.
%
%
%
%

\subsection{\label{sec:5.1}Determination of the threshold}

We investigate in this section how the spatial modulation $M(x)$
changes the bifurcation scenario from the basic  state $A=0$ of Eq.~(\ref{Foampe})
into a spatial pattern. 
In the absence of the modulation, i.e. $G=0$,
the   linear part of Eq.~(\ref{Foampe}) is solved by $A=Fe^{\lambda t+iqx}$
and this ansatz leads in the neutrally stable case ${\rm Re}(\lambda)=0$
to an expression for the neutral curve $\varepsilon_0(q)$ and
the frequency dispersion $\omega_0(q)={\rm Im}(\lambda)$,
\begin{align}
\varepsilon_0(q) &= q^2\, , & \omega_0(q) &=-bq^2 \,\, .
\end{align}
Minimizing $\varepsilon_0(q)$ determines the threshold $\varepsilon_c=0$,
the critical wave number $q_c=0$ and the frequency $\omega_c=0$. 
In the presence of $M(x)$ the bifurcation properties of
Eq.~(\ref{Foampe}) are changed,  as illustrated in the following
by a perturbation calculation for small modulation
amplitudes.

\subsubsection{\label{sec:5.1.1}Perturbation method for small amplitudes of $M(x)$}

For small forcing amplitudes $G \ll 1$ we introduce  
 a small expansion parameter $\eta \ll 1$ with $M(x)= \eta {\bar M}(x)$.
Since the  amplitude equation~(\ref{Foampe}) is of first order with respect to 
time,  the solution of its linear part has an exponential time dependence
and for small values of the modulation amplitude,  $G=\eta {\bar G}$, 
the linear solution may be expanded in powers of the modulation
strength $\eta$
\begin{eqnarray}
\label{CFAexpansion}
 A  = e^{\lambda t} \left( A_0 + \eta A_1 + \eta^2 A_2 + \cdots \right ) \, . 
\end{eqnarray}
Applying the neutral stability condition ${\rm Re}(\lambda)=0$, 
the perturbation in Eq.~(\ref{CFAexpansion}) does neither grow
nor decay, which separates the parameter regime
where the basic  state $A=0$ is stable from the range where
$A=0$ is unstable.

The threshold is shifted due to  the modulation 
and therefore, the  control parameter $\varepsilon_c$ and the frequency $\omega_c={\rm Im}(\lambda)$
are also expanded with respect to $\eta$
\begin{subequations}
\label{CFepsomegexpansion}
\begin{eqnarray}
 \varepsilon_c &=& \eps{0} + \eta \eps{1} + \eta^2 \eps{2} + \cdots \,\, , \\
 \omega_c &=&  \omega_c^{(0)} + \eta \omega_c^{(1)} + \eta^2 \omega_c^{(2)} +\cdots \,\, .
\end{eqnarray}
\end{subequations}
The expansions given in Eqs.~(\ref{CFAexpansion}) and (\ref{CFepsomegexpansion}) 
provide the following hierarchy of equations
defining the neutral stability of the basic  state $A=0$:
\begin{subequations}
\label{FoAtot}
\begin{align}
\label{FoA0e}
\eta^0: \,\,\, {\cal L}_0 A_0 &= 0 \, , \\
\label{FoA1e}
\eta^1: \,\,\, {\cal L}_0 A_1&= [   \eps{1} + i\omega_c^{(1)} +(1+is_2){\bar M} ] A_0 \, , \\
\label{FoA2e}
\eta^2: \,\,\, {\cal L}_0 A_2&=  [   \eps{1}+ i\omega_c^{(1)} +(1+is_2){\bar M} ] A_1 \nonumber \\
                                                  & +[\eps{2}+ i\omega_c^{(2)} ] A_0 \, ,
\end{align}
\end{subequations}
with the linear operator 
${\cal L}_0=\partial_t -i\omega_c^{(0)} -\eps{0}-(1+ib) \partial_x^2$.
These equations may be solved by a spatial dependence which is either harmonic
\begin{subequations}
\begin{align}
 A_0 &=F_0  \, ,   \\
 A_1 &= F_2e^{2ikx} + F_{-2}e^{-2ikx}  \, , \\
   &  \cdots  \nonumber
\intertext{or subharmonic}
A_0 &= F_1e^{ikx}  + F_{-1} e^{-ikx} \, , \\
A_1 &= F_3 e^{3ikx} + F_{-3} e^{-3ikx}  \, , \\
   &   \cdots \nonumber
\end{align}
\end{subequations}
with respect to $M(x)$.
In order to distinguish between the harmonic and subharmonic results
we introduce $\varepsilon_h, \omega_h$ for the harmonic case and
 $\varepsilon_{sh}, \omega_{sh}$ for the subharmonic one.
From a solubility condition on  the right hand side of 
Eqs.~(\ref{FoA1e}) and (\ref{FoA2e})  the corrections to $\varepsilon_c$
and $\omega_c$ may be calculated. The solution $A_1$ of Eq.~(\ref{FoA1e})
is given for the harmonic ansatz by
\begin{eqnarray}
  A_{1} = \frac{G(1+is_2) F_0} {4k^2(1+ib)} \left( e^{2ikx} + e^{-2ikx} \right) ,
\end{eqnarray}
and in the case of the subharmonic ansatz by
\begin{eqnarray}
  A_1 = \frac{G(1+is_2)} {8k^2(1+ib)} \left( F_1 e^{3ikx} +  F_{-1} e^{-3ikx}\right)  ,
\end{eqnarray}
whereas the solution $A_2$ of Eq.~(\ref{FoA2e}) is not needed explicitly
to determine the corrections $\eps{2}$ and $\omega_c^{(2)}$, respectively.
The expansions of the 
threshold $\varepsilon_c$ and the frequency $\omega_c$
up to order $O(\eta^2)$ are given
for the harmonic case by
\begin{subequations}
\label{CFoallharm}
\begin{align}
\label{CFoepshpert}
 \varepsilon_{h} &= -\eta^2 \frac{ {\bar G}^2 ( 1-s_2^2+2bs_2 )} {2k^2(1+b^2)} \, ,\\
\label{CFoomeghpert}
 \omega_{h} &= \eta^2 \frac{ {\bar G}^2  \left[b( 1-s_2^2)-2s_2 \right]} {2k^2(1+b^2)} \, ,
\end{align}
\end{subequations}
and for the subharmonic case by
\begin{subequations}
\label{CFoallsub}
\begin{align}
\label{CFoepsshpert}
\varepsilon_{sh} &= k^2 -\eta {\bar G} -\eta^2 \frac{{\bar G}^2 ( 1-s_2^2+2bs_2 )} {8k^2(1+b^2)} \, ,\\
\omega_{sh} &= bk^2 -\eta s_2{\bar G} 
              + \eta^2 \frac{ {\bar G}^2  \left[b( 1-s_2^2)-2s_2 \right]} {8k^2(1+b^2)} \, .
\end{align}
\end{subequations}
If one increases the control parameter in Eq.~(\ref{Foampe}) from below,
the lowest of the  two thresholds $\varepsilon_h$ and $\varepsilon_{sh}$
determines whether the basic state $A=0$
is unstable against spatially harmonic solutions, i.e. if $\varepsilon_h<\varepsilon_{sh}$,
or spatially subharmonic solutions,  i.e.  if $\varepsilon_{sh}<\varepsilon_{h}$.
By replacing $k \to \xi_0 k$ and ${\bar G} \to s_1 {\bar G}$ in
the expressions~(\ref{CFoallharm}) and (\ref{CFoallsub}), 
the thresholds $\varepsilon_{h,sh}$
and the frequencies $\omega_{h,sh}$
for  the amplitude equation~(\ref{CfMGLE}) follow.

\subsubsection{\label{sec:5.1.2}Numerical determination of the threshold}

Because of the periodically varying coefficient $M(x)$
in Eq.~(\ref{Foampe}), the general linear solution 
may be represented, similar as in Sec.~\ref{sec:3.3}, by a Floquet-Bloch expansion
\begin{equation}
\label{CFoAansatz}
 A(x,t) = e^{\lambda t +iqx} \sum_{j=-N}^{N} F_j e^{ij2kx}  
                      \quad  \left( {\rm with}  \,\,  0 \leq q<2k \right ) ,
\end{equation}
up to an  appropriate number $N$. Without spatial forcing
all coefficients but $F_0$ vanish.
The perturbation in Eq.~(\ref{CFoAansatz}) grows for a chosen
parameter combination if ${\rm Re}(\lambda)>0$ and it decays
if ${\rm Re}(\lambda)<0$. We are again primarily interested in the neutrally stable
case ${\rm Re}(\lambda)=0$, separating the stable from the unstable regime. 
After inserting the ansatz for $A$ into the linear part of Eq.~(\ref{Foampe}), 
the explicit $x$-dependence is removed
by  multiplying  the  equation with  $\exp{(-il2kx)}$ ($l=-N, \ldots , N$)
and integrating  with respect to $x$.
From this procedure the eigenvalue problem
\begin{eqnarray}
\label{CFoeigvalue}
 \mathcal{A} \vec{F} = \rho \vec{F} \qquad \left[ \vec{F}=(F_{-N}, \ldots, F_N) \right ] 
\end{eqnarray}
follows,  where the matrix $\mathcal{A}$ is a band matrix of width $(2N+1)$ with
the coefficients
\begin{subequations}
\label{CFomatrixele}
\begin{eqnarray}
 \mathcal{A}_{l,l} &=& (1+ib) (q+2lk)^2  \, , \\
 \mathcal{A}_{l,l-2} &=& -(1+is_2) G \,, \\
   \mathcal{A}_{l,l+2}  &=& \mathcal{A}_{l,l-2}  \, .
\end{eqnarray}
\end{subequations}
From a  solvability condition for the homogeneous 
system of equations~(\ref{CFoeigvalue}),  
\begin{eqnarray}
 {\rm det}(\mathcal{A}-\rho \mathcal{I}) = f(G, k, q \ldots) =0\, , 
\end{eqnarray}
$(\mathcal{I}$ is the unity matrix)
the eigenvalues $\rho_i$ are determined as  functions of the
parameters and they can  be sorted in ascending order with
respect to their real parts ${\rm Re}(\rho_i)$.
Keeping all parameters besides  $q$ fixed,
the neutral curve $\varepsilon_0(q)$ and the frequency
$\omega_0(q)$ can be determined
from the  eigenvalue with the lowest  real part
$\widehat{\rho}={\rm min} [{\rm Re}(\rho_i)]$.
Minimizing $\varepsilon_0(q)$ gives then the threshold $\varepsilon_c$,
the critical wave number $q_c$ and the critical frequency $\omega_c=\omega_0(q=q_c)$.
In this way,  we found  that the minimum of the neutral curve
is either given at $q_c=0$ determining the harmonic threshold $\varepsilon_h$, 
or at $q_c=k$ determining the subharmonic threshold $\varepsilon_{sh}$.
The harmonic and subharmonic contributions  of the ansatz in Eq.~(\ref{CFoAansatz})
with respect to $M(x)$ separate for the linear 
part of Eq.~(\ref{Foampe}), and 
the two thresholds $\varepsilon_{h}$ and $\varepsilon_{sh}$
may be calculated independently.

In order to obtain the analogous eigenvalue equation 
with respect to the unscaled amplitude 
equation given in Eq.~(\ref{CfMGLE}), one has to replace in Eqs.~(\ref{CFomatrixele})
the wave numbers $q$ and $k$
by $\xi_0q$ and $\xi_0k$, respectively, 
as well as the amplitude $G$ by $s_1G$. 

\subsection{\label{sec:5.2}Results}

One interesting question is the location of the boarder
separating the  parameter range where
the harmonic pattern is preferred at threshold from
that range where the subharmonic pattern is favored, whereby
the borderline is determined
by the condition $\varepsilon_h=\varepsilon_{sh}$.
In terms of our perturbational results as given in Sec.~\ref{sec:5.1.1}
this condition leads to a second order polynomial in 
the modulation amplitude $G$ with its 
two solutions 
\begin{eqnarray}
\label{CFoGpm}
  G_{\pm} &=& \frac{4k^2(1+b^2)}{3(1-s_2^2+2bs_2)} \pm \frac{2k^2(1+b^2)}{3(1-s_2^2+2bs_2)} \nonumber \\
&&\times  \sqrt{4-6\frac{1-s_2^2+2bs_2}{1+b^2}} \,\, . 
\end{eqnarray}
There is a finite range in $G$ 
where subharmonic solutions are preferred, namely
if  the following inequality 
\begin{eqnarray}
\label{CFoinequGpm}
       4 \left(1+b^2 \right) -6 \left(1-s_2^2+2bs_2 \right) >0 \, 
\end{eqnarray}
is fulfilled.
%
\begin{figure}
\includegraphics [width=0.45\textwidth] {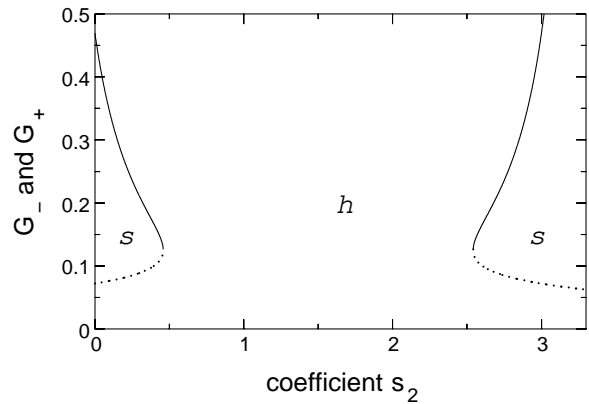} 
     \caption {                                                                                                 
       The two solutions $G_{+}$ (solid lines) and  $G_{-}$ (dotted lines), 
       as  given by Eq.~(\ref{CFoGpm}),
       are shown as a function  of the coefficient $s_2$ and 
       for the parameters $k=0.25, b=1.5$.
       The areas with  pure harmonic and subharmonic 
       solutions at threshold are marked by h and s, respectively.
                   }                         
\label{fig:8}                                                                    
\end{figure}                                                                                                    
%
%
The harmonic threshold is the lowest one 
for modulation amplitudes $G<G_{-}$ and $G>G_{+}$.
In the finite range $G_{-}<G<G_{+}$ the subharmonic threshold 
$\varepsilon_{sh}$
drops 
below the harmonic one.
The two amplitudes $G_{+}$ and $G_{-}$
according to the formula~(\ref{CFoGpm}) are plotted in Fig.~\ref{fig:8}
as functions of the coefficient $s_2$. 
The ranges  in which subharmonic or harmonic solutions 
are preferred  are marked by s and h, respectively. 

In the parameter range $s_2>b+\sqrt{1+b^2}$ and $s_2<b-\sqrt{1+b^2}$
the harmonic threshold has a positive curvature as a function 
of $G$, $\partial^2\varepsilon_h/\partial G^2>0$,
which can be readily verified from Eq.~(\ref{CFoepshpert}).
Consequently, at small values of $G$
the threshold $\varepsilon_h$ is shifted upwards
while the subharmonic 
threshold is dominated by the linear decrease
$\varepsilon_{sh}\propto k^2-G$.
These trends of the two thresholds 
apparently promote  the appearance of subharmonic patterns.

%
%
\begin{figure*} 
 \includegraphics [width=7.5cm] {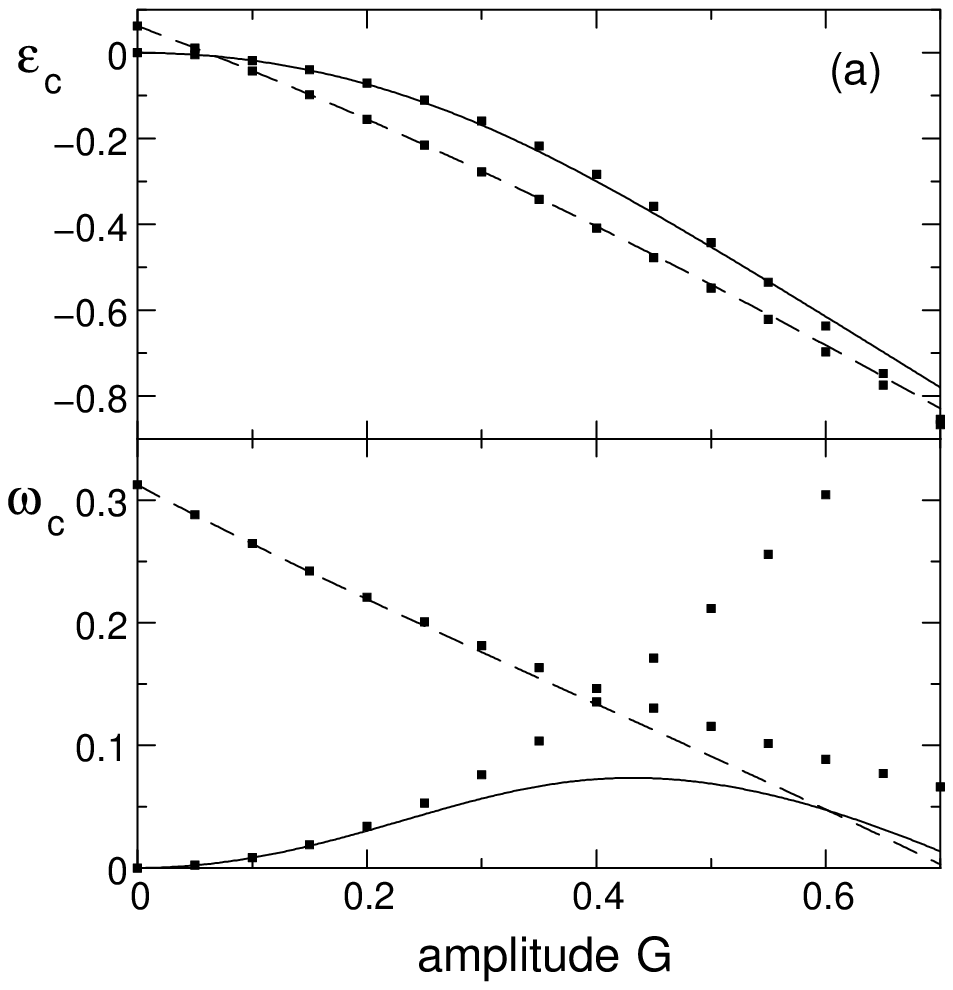} 
\hspace*{0.4cm}
 \includegraphics [width=7.5cm] {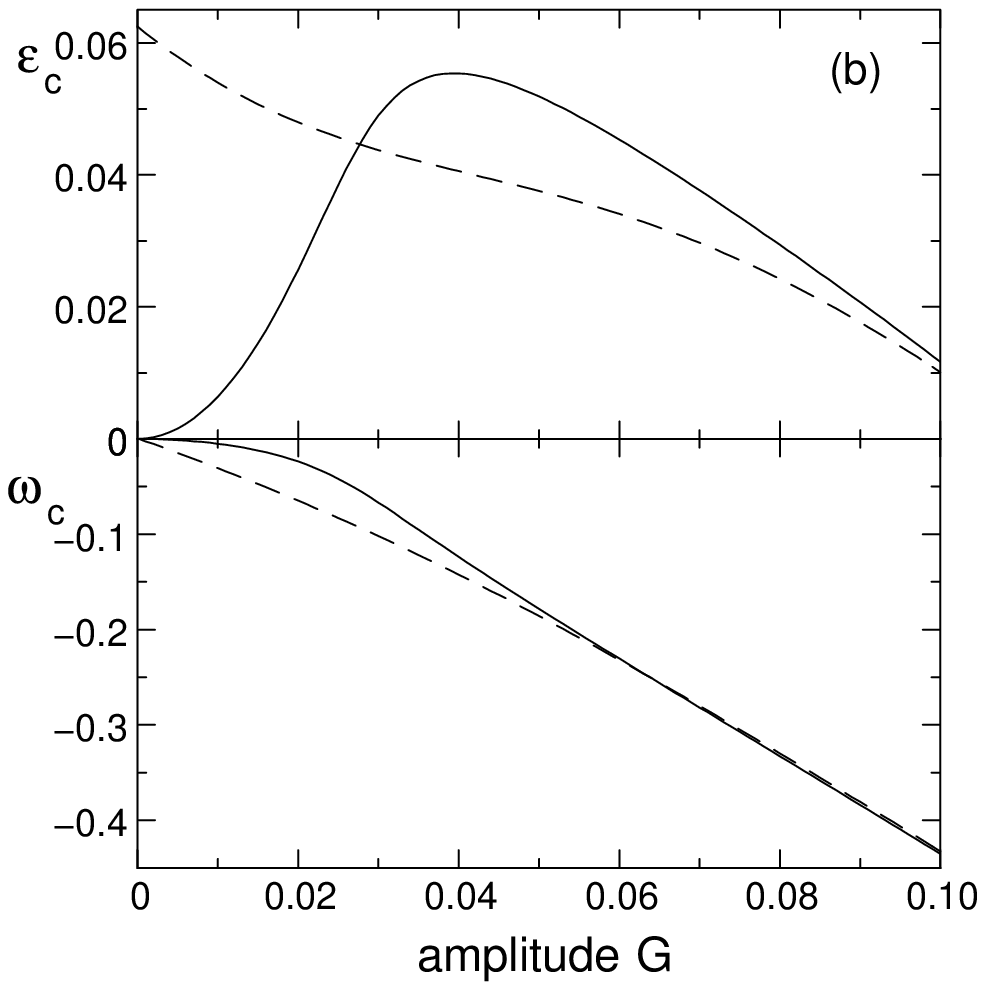} 
     \caption {                                                               
       The harmonic (solid lines) and subharmonic (dashed lines) threshold as well as
       the corresponding frequencies are plotted as functions of the forcing amplitude $G$
       for two sets of parameters, in part (a) for $k=0.25,~ b=5, ~s_2=0.5$
       and in part (b) for $k=0.25,~ b=0.01, ~s_2=3$. In order to
       generate the numerical
       results 
       $N=32$ modes in the Floquet expansion (\ref{CFoAansatz}) have been used.
       Subharmonic patterns occur in the $G$-range where $\varepsilon_{sh}<\varepsilon_h$.
       The second point of intersection between the harmonic
       and subharmonic branch is located at $G \simeq 1.53$ in part (a) and 
       at $G \simeq 0.12$ in part (b). For comparison, 
       the symbols in part (a) are obtained by the analytical expressions 
       given in Eqs.~(\ref{CFoallharm}) and~(\ref{CFoallsub}).
                  }                                                                                          
\label{fig:9}
\end{figure*}                                                                                                 
%
%

For arbitrary values of the 
modulation amplitude $G$ and wave number $k$
the linear stability of the basic  state $A=0$ of Eq.~(\ref{Foampe})
has to be determined  numerically
by solving the eigenvalue problem (\ref{CFoeigvalue}).
The harmonic threshold $\varepsilon_h$ and the subharmonic
threshold $\varepsilon_{sh}$ are calculated separately 
and some results for them as well 
as for the corresponding critical frequencies 
$\omega_h$ and $\omega_{sh}$, respectively, 
are shown in Fig.~\ref{fig:9} 
for two sets of parameters.
The harmonic branches $\varepsilon_h, \omega_h$ 
(solid lines) and subharmonic branches $\varepsilon_{sh}, \omega_{sh}$ (dashed lines)
have been calculated for $N=32$ Fourier modes,
which has been proven  to be a reasonable approximation.
The symbols in part (a) indicate the results of the perturbation 
calculation as given in Eqs.~(\ref{CFoallharm}) and~(\ref{CFoallsub}),
which are  in good agreement with the numerical results for small 
forcing amplitudes $G$.
The points of intersection between the branches $\varepsilon_h$ and 
$\varepsilon_{sh}$ are roughly given by $G_{-} \simeq 0.068$, $G_{+} \simeq 1.53$
in part (a) and by $G_{-} \simeq 0.028$, $G_{+} \simeq 0.12$ in part (b).
For comparison, the perturbation calculation yields
$G_{-} \simeq 0.068$, $G_{+} \simeq 0.68$
for the parameters in part (a) and 
$G_{-} \simeq -0.048$, $G_{+} \simeq 0.027$ for the ones in part (b).
As mentioned  above, the sign of the curvature 
$\partial^2\varepsilon_h/\partial G^2$
at small amplitudes  $G$ 
may be changed by varying the coefficients $b$ and $s_2$, which
can be recognized by comparing the course of $\varepsilon_h$
in part (a) given for $(b,s_2)=(5,0.5)$
and in part (b) given for $(b,s_2)=(0.01,3)$.

Harmonic solutions are preferred 
for large modulation wave numbers $k$, 
and in the limiting case $k \to \infty$, 
the  harmonic threshold approaches $\varepsilon_h=0$ 
while the subharmonic threshold diverges in this limit,
being  in agreement with the expressions
given in Eqs.~(\ref{CFoepshpert}) 
and~(\ref{CFoepsshpert}). 
In  the opposite limit, i.e.  $k\to 0$, 
the thresholds
$\varepsilon_h$ and $\varepsilon_{sh}$ 
tend to $\varepsilon_{h,sh}=-2G$, 
which has to be computed numerically.
We remind the reader  that the thresholds
$\phi_{0c}^h(k)$ and $\phi_{0c}^{sh}(k)$,  
as obtained for the Lengyel-Epstein model,
exhibit qualitatively the same behavior
as a function of $k$, cf.~Fig.~\ref{fig:5}(a).

\begin{figure} 
 \includegraphics [width=0.45\textwidth] {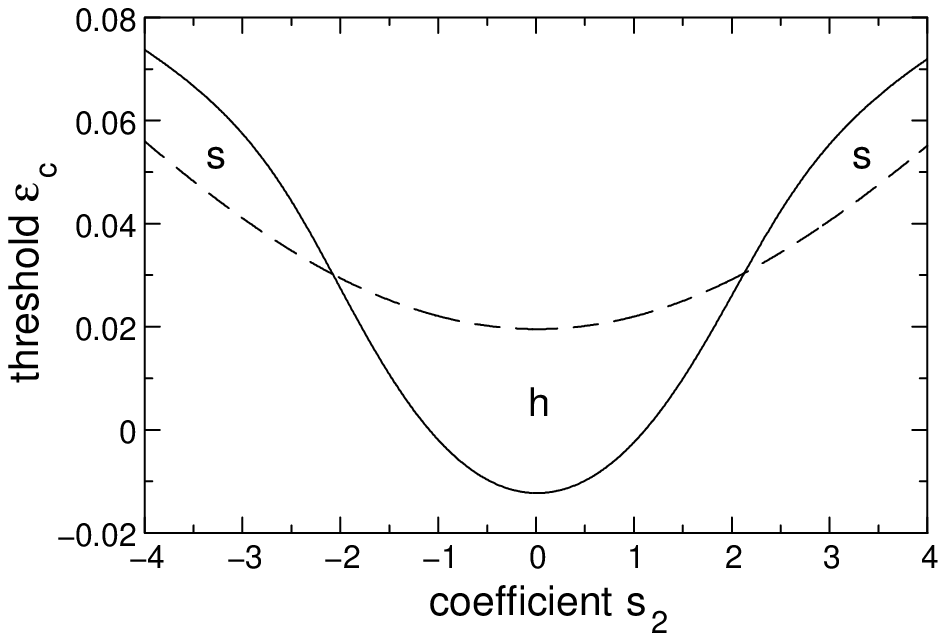} \\
 \includegraphics [width=0.45\textwidth] {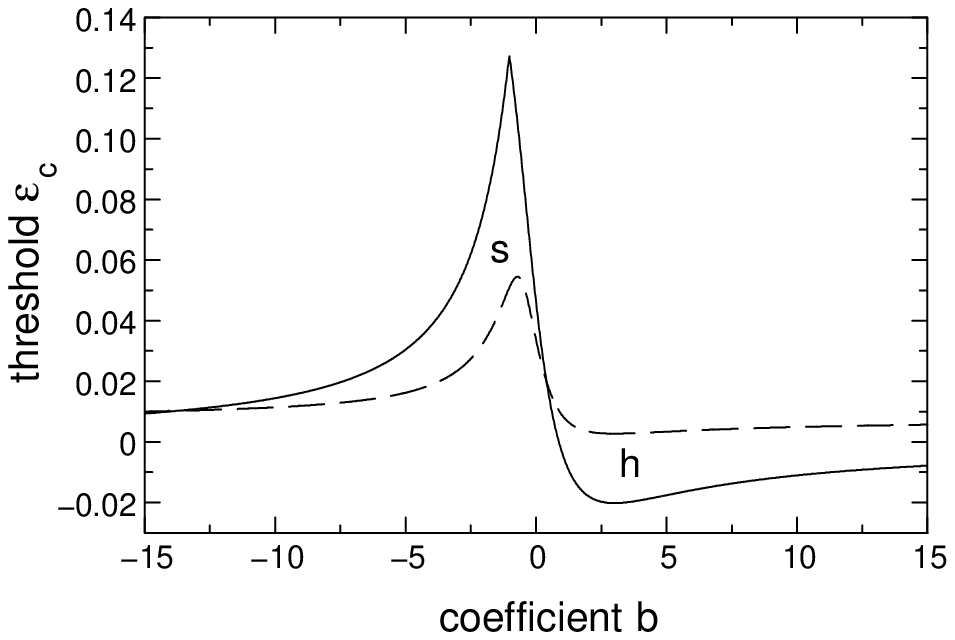} 
     \caption {                                                                                              
       The dependence of the harmonic (solid line)  and subharmonic
       (dashed line) threshold on the coefficient $s_2$ is shown in the top part
       for  a set of parameters given by $G=0.04,~ k=0.25,~ b=0.01$.
       The bottom part  shows both  thresholds 
       as function of the coefficient $b$ and
       for the parameters $G=0.05,~ k=0.24,~ s_2=3$.
       The range where harmonic solutions are preferred is
        marked by h
       and that of  subharmonic ones by s. 
                  }                                                                                          
\label{fig:10}
\end{figure}

In Fig.~\ref{fig:10} the harmonic threshold $\varepsilon_h$ (solid line)
and the subharmonic threshold $\varepsilon_{sh}$ (dashed line)
are plotted as functions of the linear coefficient $s_2$ (top)
and $b$ (bottom).
For large values of the modulus $|s_2|$  the subharmonic
threshold drops below the harmonic one 
and spatially subharmonic solutions appear at threshold of the Hopf
bifurcation. 
Regions solely supporting harmonic or subharmonic
patterns are found in the range s and h, respectively.
The points of intersection between both  thresholds  are roughly given by
$s_2 \simeq \pm 2$. 
For comparison, $s_2\simeq \pm 1.84$ is  obtained
from  the perturbation calculation.
According to the bottom part of the figure,
subharmonic solutions are preferred in a finite range of the parameter $b$
similar as the $G$-range in  Fig.~\ref{fig:9}.
Again, within the regions marked
by s and h only subharmonic or harmonic patterns
are found.

From the inequality given in Eq.~(\ref{CFoinequGpm})
one may easily deduce the following limiting cases. For
$b=0$ subharmonic solutions are favored in the
range $ s_2^2>1/3$ and for $s_2=0$ in the
range $  b^2>1/2$,
while $b=s_2=0$ seems to lead to a contradiction.  
Therefore, one of the two  linear coefficients $b$ and $s_2$
is necessary for the occurrence of spatially subharmonic patterns.

%
%
%
%

\subsubsection{Comparison with the Lengyel-Epstein model}
\label{sec:5.3.1}

In order to be able
to compare directly the results for the harmonic and subharmonic
thresholds with
those obtained from the basic equations~(\ref{Cfepstein}),  we have
to consider the linear part of 
the unscaled amplitude equation~(\ref{CfMGLE}), which
reflects the typical length scale, $\xi_0$, time scale, $\tau_0$, 
and amplitude scale, $s_1$,
of the Lengyel-Epstein model. 
In this context one may raise the question under which conditions are
the thresholds $\varepsilon_h, \varepsilon_{sh}$  determined
by the amplitude-equation  approach, 
and the thresholds $\phi_{0c}^h, \phi_{0c}^{sh}$
determined  by the  Lengyel-Epstein model,
in (good) agreement?
Such a comparison of the thresholds, 
e.g.  as a function of the modulation amplitude $G$,
gives also some insights into  the validity range of the amplitude equation
which is a priori unknown.

For this comparison,   we have to replace in the eigenvalue 
problem in Eq.~(\ref{CFoeigvalue}) the wave numbers $q$ and $k$
by $\xi_0 q$ and $\xi_0 k$ as well as the amplitude
$G$ by $s_1G$.
Furthermore,  the values of the linear coefficients
$\tau_0, c_1, \xi_0, b, s_1$ and  $s_2$ are determined by the 
parameters of the Lengyel-Epstein model as shown, 
for instance, in Fig.~\ref{fig:7}.

\begin{figure} 
 \includegraphics [width=0.48\textwidth] {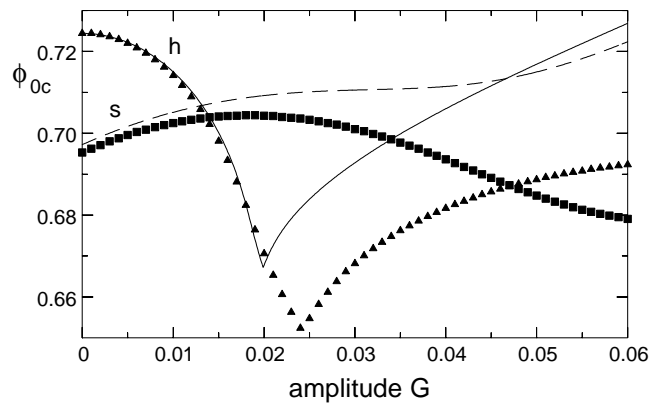}
     \caption {                                                                                                 
       \label{fFcomp_LEamp}
       Comparison between the harmonic (h) and subharmonic (s)
       thresholds,   which have been calculated  for the Lengyel-Epstein model (symbols) 
       by solving the eigenvalue equation~(\ref{FoLEeigvalue})
       and for the amplitude equation (lines) by
       solving the eigenvalue equation~(\ref{CFoeigvalue}). 
       For the latter one,  we have replaced the wave numbers $q$ and $k$
       by $\xi_0q$ and $\xi_0k$ as well as the amplitude $G$ by $s_1G$. 
       The modulation wave number is $k=0.05$. 
       The other parameters are given by $a=10, c=0.55, \sigma=5, d=1$ leading
       to the  coefficients of the amplitude equation~(\ref{CfMGLE}) 
       as given in Fig.~\ref{fig:7} (dotted line).
                  }
\label{fig:11}                                                                                             
\end{figure}

Figure~\ref{fFcomp_LEamp} shows the  harmonic threshold (triangles)
and the subharmonic one (squares)  as given for the Lengyel-Epstein model
as well as the related   thresholds obtained from the amplitude equation (lines).
The latter ones are calculated  with respect to the illumination rate 
by using the formula
$ \phi_{0c} = \phi_{0c}(G=0) ( 1-\varepsilon_c )$. 
According to this figure, one finds a qualitatively similar 
behavior of the associated  thresholds. 
For comparison, the relative error between the harmonic thresholds
is roughly given by $5 \%$ and between  the subharmonic 
thresholds by $6 \%$   calculated for $G=0.06$.
These  deviations become  smaller 
for  decreasing values of the wave numbers $k$ or by decreasing the 
modulation amplitude $G$.
Moreover, 
the finite $G$-range in which the subharmonic solution has the 
higher  threshold is also in good agreement with the range given
by the amplitude approach.

%
%
%
%

Another interesting comparison  between both models
is  the change of the sign of the curvature of the harmonic 
threshold,  which has been found   for the amplitude equation, cf.~Eq.~(\ref{CFoepshpert}).
This significant behavior  of the threshold
has also been observed    for the Lengyel-Epstein model
in a parameter regime determined  by  the amplitude equation. 
If the curvature of $\phi_{0c}^h(G)$ is positive 
the Hopf bifurcation 
also occurs for larger illuminations $\phi_0$,
being in contrast to the situation shown in
Fig.~\ref{fig:4}(a) or Fig.~\ref{fig:11}.

\section{\label{sec:6}Summary and conclusions}

We have investigated the effects of a spatially periodic modulated
control parameter  on 
an oscillating chemical reaction described 
by the Lengyel-Epstein model.
This reaction exhibits a supercritical Hopf bifurcation
and it provides due to its photosensitivity
a simple approach to study the response
of the reaction with respect to a spatially modulated
illumination.

We find that in the range of intermediate values of the
modulation amplitude $G$ the bifurcation to
the oscillatory chemical reaction is subharmonic
with respect to the external modulation whereas
for small and large modulation amplitudes the
bifurcation is harmonic. Beyond the bifurcation
point the subharmonic solution is preferred in
a finite range of the control parameter before 
a transition to the harmonic pattern takes place at
larger values.  The related results of the stability
calculations of the basic state with respect to
small perturbations are summarized in Figs.~\ref{fig:4}
and~\ref{fig:5}.

Close to the threshold of the Hopf bifurcation,
an   amplitude equation is presented
which is an extension of the well-known
complex Ginzburg-Landau equation  for 
spatially homogeneous, oscillatory bifurcations. 
This equation is generic for oscillatory systems
near threshold underlying  a spatially varying
control parameter and  it may therefore also be found
for other systems having the same symmetry properties
as the considered Lengyel-Epstein model.

The stability limit   of the basic  state of the amplitude equation 
has been determined by a perturbation calculation and by
solving the general linear problem numerically.
Good agreement of the two approaches is found for small
forcing amplitudes. 
It has been shown that intermediate forcing
amplitudes  lead also to
subharmonic solutions,  while weak and strong forcing 
amplitudes favor harmonic solutions.
A rough estimation of the validity range of the amplitude
equation has been  given by comparing the harmonic and subharmonic thresholds
with those obtained for the Lengyel-Epstein model itself. 
We have found that the amplitude equation describes 
the linear properties  of the Lengyel-Epstein model
to a great extent for long-wavelength modulations 
and small forcing amplitudes.

According to our results we expect in experiments on chemical
reactions, which are described by the 
Lengyel-Epstein model, also a transition to
subharmonic patterns induced by a spatially periodic illumination.

Instead of considering stationary forcing, 
the effect of spatiotemporal forcing
on an  oscillating chemical system may also be investigated.
Here, the forcing has the form of a traveling wave
similar as introduced recently to study
the effects of spatiotemporal forcing on 
stationary Turing patterns~\cite{Sagues:2004.1,Sagues:2003.1}
as well as on oblique stripe patterns in anisotropic 
systems~\cite{Schuler:2004.1}.
This special type of forcing 
breaks a further symmetry of the system,
the  reflection symmetry,
which may induce a more complex 
spatiotemporal behavior   to which forthcoming
work is devoted.

\appendix*
\section{\label{appendix}Scheme for the derivation of the amplitude equation}

For the derivation of the amplitude
equation~(\ref{CfMGLE}),  a small reduced control parameter 
is introduced,
\begin{equation}
\varepsilon = \frac{\phi_{0c}-\phi_0}{\phi_{0c}} \,\, ,
\end{equation}
which is a measure for the distance from the threshold
of the Hopf bifurcation. 
At threshold the linear solutions of the Lengyel-Epstein model
may be written as
\begin{equation}
\label{CFou1uthrsh}
 \mathbf{u}_1 =  A \mathbf{u}_e \, e^{i\omega_c t} + {\rm c.c.} \, ,
\end{equation}
where $A$ describes the common amplitude of the two fields $u_1, v_1$
and the eigenvector  $\mathbf{u}_e=(1, E_0)^T$ 
describes  their amplitude ratio,  cf.~Eq.~(\ref{Cfstabu1v1}).

For the perturbation expansion we introduce 
slow time and space variables~\cite{CrossHo}
\begin{align}
   X&= \varepsilon^{1/2}x \, ,& T_1 &= \varepsilon^{1/2} t\, , & T&=\varepsilon t \, 
\end{align}
and  the solution $\mathbf{u}_1$ 
may be written as a product of a slowly varying amplitude and a fast oscillating
exponential function
\begin{equation}
\label{CFou1ansatz}
 \mathbf{u}_1 =  A(X,T,T_1)  \mathbf{u}_e \, e^{i\omega_c t} + {\rm c.c.} \, .
\end{equation}
According to the  chain rule one may  replace time and space derivatives by 
the following expressions
\begin{align}
 \partial_t & \to \partial _t +  \varepsilon^{1/2} \partial_{T_1} +  \varepsilon \partial_T \, , &
\partial_x  &\to \partial_x+  \varepsilon^{1/2} \partial_X \, .
\end{align}
Note that $\partial_t$ and $\partial_x$ on the right-hand side
only act on the rapid dependences. 
We further assume small modulation amplitudes
$M(x)=\varepsilon {\bar M}(x)$ with ${\bar M}(x) \propto O(1)$
and wave numbers $k = \varepsilon^{1/2} {\bar k}$ 
with ${\bar k} \propto O(1)$.
An attribute of the $\pm$ symmetry of a supercritical oscillatory bifurcation
is the power law for the oscillation amplitude $A\sim \sqrt{\varepsilon}$.
Accordingly, the solutions of the basic equation~(\ref{Cfepstein2}) 
are expanded near threshold with respect to powers of 
$\varepsilon^{1/2}$,
\begin{equation}
 \mathbf{u} = \mathbf{u}_0 +  \varepsilon^{1/2} \mathbf{u}_1 +  \varepsilon \mathbf{u}_2 
                    +  \varepsilon^{3/2}\mathbf{u}_3 + \cdots \,\, .
\end{equation}
The components of $\mathbf{u}_0$ describe the stationary solutions
as given in Eq.~(\ref{Cfepsteinstat})  and the components of 
$\mathbf{u}_1=A_1 \mathbf{u}_e\exp{(i\omega_c t)}+{\rm c.c.}$ 
provide the linear oscillatory solutions at threshold as discussed 
in Sec.~\ref{sec:3.1}.
Note that the amplitude $A_1$ and the amplitude $A$ in Eq.~(\ref{CFou1ansatz})
only differ by a factor of $\varepsilon^{1/2}$.
The expansions for $\mathcal{L}, \mathbf{N}$ and $\mathbf{V}$
in Eq.~(\ref{Cfepstein2}) are given by
\begin{subequations}
\label{CFolinexpansion}
\begin{eqnarray}
  \mathcal{ L} &=&  \mathcal{ L}_0 + \varepsilon^{1/2} \mathcal{ L}_1  +
                                    \varepsilon \mathcal{ L}_2  \,\, ,\\
\label{CFolinexpansion_b}
   \mathbf{N} &=&  
\mathbf{N}_0  + \mathbf{L} + 
                          \varepsilon \mathbf{N}_2 + \varepsilon^{3/2} \mathbf{N}_3 + \cdots  \,\, ,\\
{\rm with} \quad    \mathbf{L} &=& \varepsilon^{1/2}\mathcal{ M}_0 \mathbf{u}_1 + 
                              \varepsilon \mathcal{ M}_0 \mathbf{u}_2 + 
                              \varepsilon^{3/2}\mathcal{ M}_0 \mathbf{u}_3 + \cdots \nonumber \,\, ,\\
\label{CFolinexpansion_c}
   \mathbf{V} &=&  \mathbf{V}_0  + \varepsilon  \mathbf{V}_2   \, \, ,
\end{eqnarray}
\end{subequations}
and with the explicit expressions of the
linear operators ${\mathcal L}_0, {\mathcal L}_1$ as well as  ${\mathcal L}_2$: 
\begin{align}
{\mathcal L}_0 &=
    \begin{pmatrix}
                  \partial_t +c & 0 \\
                    -\sigma c       & \partial_t 
    \end{pmatrix} , \nonumber \\
{\mathcal L}_1 &=
    \begin{pmatrix}
                  \partial_{T_1} -2 \partial_x \partial_X  & 0 \\
                                0            & \partial_{T_1}  - 2 \sigma d \partial_x \partial_X
    \end{pmatrix} , \nonumber \\
{\mathcal L}_2 &=
    \begin{pmatrix}
                  \partial_T -\partial_X^2 & 0 \\
                    0       & \partial_T - \sigma d \partial_X^2
    \end{pmatrix} .
\end{align}
The nonlinear  vectors in the expansions~(\ref{CFolinexpansion_b}) 
and~(\ref{CFolinexpansion_c})  are given by
\begin{eqnarray}
{\mathbf N}_0 &=& \frac{u_0v_0}{1+u_0^2}
    \begin{pmatrix}
                   - 4 \\  -\sigma 
    \end{pmatrix} , \nonumber \\
{\mathbf N}_2 &=&- 
    \begin{pmatrix}
                4 \\
                 \sigma
    \end{pmatrix}\left(  E_1  u_1 v_1  + E_2 u_1^2  \right) , \nonumber \\
\label{CfNlineps3}
{\mathbf N}_3 &=&- 
    \begin{pmatrix}
                4 \\
                \sigma
    \end{pmatrix} {\cal N}_3~ ,\qquad
 {\rm with} \nonumber \\
 {\cal N}_3& =&  
        \left[ R_1 (u_1 v_2  +  u_2 v_1) +R_2 u_1 u_2+ R_3 v_1u_1^2+R_4u_1^3  \right ]
\nonumber
\end{eqnarray}
and 
\begin{eqnarray}
{\mathbf V}_0 =
    \begin{pmatrix}
                   a - \phi_{0c}  \\ \sigma \phi_{0c} 
    \end{pmatrix} , \quad 
{\mathbf V}_2 = 
    \begin{pmatrix}
                \phi_{0c} - {\bar M} \\ -\sigma \phi_{0c} +\sigma {\bar M}
    \end{pmatrix} ,
 \end{eqnarray}
where the abbreviations 
\begin{align}
E_1 &= \frac{1-u_0^4}{\left( 1+u_0^2 \right)^3}\, , & 
E_2 &= \frac{v_0\left(u_0^3-3u_0\right) } {\left( 1+u_0^2 \right)^3} \, , \nonumber \\
R_1 &= \frac{1+u_0^2-u_0^4-u_0^6}{\left( 1+u_0^2 \right)^4} \, , &
R_2 &= \frac{-2u_0v_0\left( 3+2u_0^2-u_0^4\right)}{\left( 1+u_0^2 \right)^4} \, ,\\
R_3 &=\frac{u_0^5-3u_0-2u_0^3}{\left( 1+u_0^2 \right)^4} \, ,&
R_4 &= \frac{6u_0^2v_0-u_0^4v_0-v_0}{\left( 1+u_0^2 \right)^4} \, ,\nonumber
\end{align}
have been introduced.
Inserting all expressions  into Eq.~(\ref{Cfepstein2}) yields at successive orders of 
$\varepsilon^{1/2}$
\begin{subequations}
\begin{align}
\label{CFoexpeps0}
\varepsilon^0: \,\,\, {\mathcal L}_0 \mathbf{u}_0  &= \mathbf{N}_0 + \mathbf{V}_0 \, ,\\
\label{CFoexpeps1}
\varepsilon^{1/2}: \,\,\, {\mathcal L}_0 \mathbf{u}_1 &= {\mathcal M}_0 \mathbf{u}_1 \, ,\\
\label{CFoexpeps2}
\varepsilon^1:  \,\,\, {\mathcal L}_0 \mathbf{u}_2 &=  {\mathcal M}_0 \mathbf{u}_2  
                - {\mathcal L}_1 \mathbf{u}_1 + \mathbf{N}_2 + \mathbf{V}_2 \, ,\\
\label{CFoexpeps3}
\varepsilon^{3/2}: \,\,\, {\mathcal L}_0 \mathbf{u}_3 &=  {\mathcal M}_0 \mathbf{u}_3 + 
                            \mathbf{N}_3 -  {\mathcal L}_2 \mathbf{u}_1 \, , \\
                                &  \cdots  \,\, . \nonumber 
\end{align}
\end{subequations}
Equation~(\ref{CFoexpeps0}) determines the basic state
as given by Eq.~(\ref{Cfepsteinstat}) 
while Eq.~(\ref{CFoexpeps1}) reproduces the threshold condition~(\ref{CfdeterA})
with the  solution known already from Eq.~(\ref{Cfeigenvalues}).
Since $\left( {\mathcal L}_0-{\mathcal M}_0 \right) \mathbf{u}_1 = \mathbf{0}$
holds due to Eq.~(\ref{CFoexpeps1}), 
the contribution ${\mathcal L}_1 \mathbf{u}_1$ on the right-hand side
of Eq.~(\ref{CFoexpeps2}) has to vanish leading to the 
constraint $\partial_{T_1} A_1 =0$. 
In order to solve the inhomogeneous equation~(\ref{CFoexpeps2}),  
we choose the following ansatz:
\begin{eqnarray}
 \mathbf{u}_2 = \begin{pmatrix} u_2 \\ v_2 \end{pmatrix} = 
      \begin{pmatrix}
                    A_0 \\
                    B_0 
    \end{pmatrix} +
      \begin{pmatrix}
                    A_2 \\ 
                    B_2 
    \end{pmatrix}  e^{2i\omega_c t}  + {\rm c.c.} \,\, .
\end{eqnarray}
For the four 
unknown amplitudes we obtain
\begin{eqnarray}
\label{CFoA-B}
A_0 &=& \frac{\phi_{0c}-{\bar M}}{c} , \nonumber \\
B_0 &=& -\frac{C_1\left( \phi_{0c}-{\bar M}\right )}{C_2\, c} -\frac{
              E_1 \left( E_0 + E_0^{\ast} \right) + 2E_2 } {C_2} \left| A_1\right |^2  ,\nonumber \\
A_2 &=& \frac{-8i\omega_c \left( E_1E_0+E_2\right )}
             {5\sigma c C_2 -4\omega_c^2 + 2i\omega_c \left( c+4C_1+\sigma C_2  \right )} \, A_1^2
                 ,\nonumber \\
B_2 &=& \frac{\sigma\left(5c+2i\omega_c \right) \left( E_1E_0+E_2\right ) }
    {5\sigma c C_2 -4\omega_c^2 + 2i\omega_c \left( c+4C_1+\sigma C_2  \right )} \, A_1^2  . 
\end{eqnarray}
After inserting $\mathbf{u}_1$ and $\mathbf{u}_2$ into Eq.~(\ref{CFoexpeps3}),
there is no need to solve this equation explicitly.
Projecting  instead the whole equation onto  $\mathbf{u}_1^{\dagger}$,
where $\mathbf{u}_1^{\dag}$ is the solution to the adjoint equation
of (\ref{CFoexpeps1}), the inhomogeneity on 
the right-hand side of Eq.~(\ref{CFoexpeps3}) yields a solubility condition.
By rescaling  back to the original units of the space and time variables
and also to the amplitudes $M=\varepsilon{\bar M}$ and $A=\varepsilon^{1/2}A_1$,
this solvability condition provides  the final form of the amplitude equation for $A$:
\begin{eqnarray}
 \tau_0 \partial_t A &=& \varepsilon \left( 1+ic_1 \right) A+ \xi_0^2 \left( 1+ib \right) \partial_x^2 A
             \\        && + s_1 \left(1+is_2 \right) M A 
                  -g\left(1+ic_2\right) \mid A \mid^2 A \, \, . \nonumber  
\end{eqnarray}
All  coefficients are given in terms 
of the parameters of the basic equations~(\ref{Cfepstein})
and we have plotted them in Fig.~\ref{fig:7}
as functions of the parameter $c$ and for three different values of $a$.


\end{document}